\documentclass[letterpaper,aps,prc,superscriptaddress,showpacs,nofootinbib,floatfix,twocolumn]{revtex4}

\usepackage{graphicx} 
\graphicspath{%
  {./figures/}%
}

\usepackage{xspace}	

\newcommand{\rdau}{\mbox{$R_{d\rm{Au}}$}\xspace}
\newcommand{\pp}{\mbox{$p$$+$$p$}\xspace}
\newcommand{\dau}{\mbox{$d$+Au}\xspace}  
 
\newcommand{\pbpb}{\mbox{Pb$+$Pb}\xspace} 
\newcommand{\uall}{\mbox{$\Upsilon(1S$$+$$2S$$+$$3S)$}\xspace}
\newcommand{\utt}{\mbox{$\Upsilon(2S$$+$$3S)$}\xspace}
\newcommand{\uone}{\mbox{$\Upsilon(1S)$}\xspace}
\newcommand{\utwo}{\mbox{$\Upsilon(2S)$}\xspace}
\newcommand{\uthree}{\mbox{$\Upsilon(3S)$}\xspace}
\newcommand{\ups}{\mbox{$\Upsilon$}\xspace}
\newcommand{\upss}{\mbox{$\Upsilon$s}\xspace}
\newcommand{\sqs}{\mbox{$\sqrt{s}$}\xspace}
\newcommand{\sqsn}{\mbox{$\sqrt{s_{_{NN}}}$}\xspace}

\begin{document}



\title{ $\Upsilon(1S$$+$$2S$$+$$3S)$ production  in $d+$Au and $p$$+$$p$ collisions at 
$\sqrt{s_{_{NN}}}$=200 GeV and cold-nuclear-matter effects}

\newcommand{\abilene}{Abilene Christian University, Abilene, Texas 79699, USA}
\newcommand{\acadsin}{Institute of Physics, Academia Sinica, Taipei 11529, Taiwan}
\newcommand{\banaras}{Department of Physics, Banaras Hindu University, Varanasi 221005, India}
\newcommand{\barc}{Bhabha Atomic Research Centre, Bombay 400 085, India}
\newcommand{\bnlcoll}{Collider-Accelerator Department, Brookhaven National Laboratory, Upton, New York 11973-5000, USA}
\newcommand{\bnlphys}{Physics Department, Brookhaven National Laboratory, Upton, New York 11973-5000, USA}
\newcommand{\caucr}{University of California - Riverside, Riverside, California 92521, USA}
\newcommand{\charlesczech}{Charles University, Ovocn\'{y} trh 5, Praha 1, 116 36, Prague, Czech Republic}
\newcommand{\chonbuk}{Chonbuk National University, Jeonju, 561-756, Korea}
\newcommand{\ciae}{Science and Technology on Nuclear Data Laboratory, China Institute of Atomic Energy, Beijing 102413, P.~R.~China}
\newcommand{\cns}{Center for Nuclear Study, Graduate School of Science, University of Tokyo, 7-3-1 Hongo, Bunkyo, Tokyo 113-0033, Japan}
\newcommand{\colorado}{University of Colorado, Boulder, Colorado 80309, USA}
\newcommand{\columbia}{Columbia University, New York, New York 10027 and Nevis Laboratories, Irvington, New York 10533, USA}
\newcommand{\czechtech}{Czech Technical University, Zikova 4, 166 36 Prague 6, Czech Republic}
\newcommand{\dapnia}{Dapnia, CEA Saclay, F-91191, Gif-sur-Yvette, France}
\newcommand{\debrecen}{Debrecen University, H-4010 Debrecen, Egyetem t{\'e}r 1, Hungary}
\newcommand{\elte}{ELTE, E{\"o}tv{\"o}s Lor{\'a}nd University, H - 1117 Budapest, P{\'a}zm{\'a}ny P. s. 1/A, Hungary}
\newcommand{\ewha}{Ewha Womans University, Seoul 120-750, Korea}
\newcommand{\fit}{Florida Institute of Technology, Melbourne, Florida 32901, USA}
\newcommand{\fsu}{Florida State University, Tallahassee, Florida 32306, USA}
\newcommand{\gsu}{Georgia State University, Atlanta, Georgia 30303, USA}
\newcommand{\hanyang}{Hanyang University, Seoul 133-792, Korea}
\newcommand{\hiroshima}{Hiroshima University, Kagamiyama, Higashi-Hiroshima 739-8526, Japan}
\newcommand{\ihepprot}{IHEP Protvino, State Research Center of Russian Federation, Institute for High Energy Physics, Protvino, 142281, Russia}
\newcommand{\illuiuc}{University of Illinois at Urbana-Champaign, Urbana, Illinois 61801, USA}
\newcommand{\inrras}{Institute for Nuclear Research of the Russian Academy of Sciences, prospekt 60-letiya Oktyabrya 7a, Moscow 117312, Russia}
\newcommand{\instpasczech}{Institute of Physics, Academy of Sciences of the Czech Republic, Na Slovance 2, 182 21 Prague 8, Czech Republic}
\newcommand{\isu}{Iowa State University, Ames, Iowa 50011, USA}
\newcommand{\jinrdubna}{Joint Institute for Nuclear Research, 141980 Dubna, Moscow Region, Russia}
\newcommand{\jyvaskyla}{Helsinki Institute of Physics and University of Jyv{\"a}skyl{\"a}, P.O.Box 35, FI-40014 Jyv{\"a}skyl{\"a}, Finland}
\newcommand{\kek}{KEK, High Energy Accelerator Research Organization, Tsukuba, Ibaraki 305-0801, Japan}
\newcommand{\korea}{Korea University, Seoul, 136-701, Korea}
\newcommand{\kurchatov}{Russian Research Center ``Kurchatov Institute", Moscow, 123098 Russia}
\newcommand{\kyoto}{Kyoto University, Kyoto 606-8502, Japan}
\newcommand{\labllr}{Laboratoire Leprince-Ringuet, Ecole Polytechnique, CNRS-IN2P3, Route de Saclay, F-91128, Palaiseau, France}
\newcommand{\lawllnl}{Lawrence Livermore National Laboratory, Livermore, California 94550, USA}
\newcommand{\losalamos}{Los Alamos National Laboratory, Los Alamos, New Mexico 87545, USA}
\newcommand{\lpc}{LPC, Universit{\'e} Blaise Pascal, CNRS-IN2P3, Clermont-Fd, 63177 Aubiere Cedex, France}
\newcommand{\lund}{Department of Physics, Lund University, Box 118, SE-221 00 Lund, Sweden}
\newcommand{\maryland}{University of Maryland, College Park, Maryland 20742, USA}
\newcommand{\mass}{Department of Physics, University of Massachusetts, Amherst, Massachusetts 01003-9337, USA }
\newcommand{\muenster}{Institut fur Kernphysik, University of Muenster, D-48149 Muenster, Germany}
\newcommand{\muhlenberg}{Muhlenberg College, Allentown, Pennsylvania 18104-5586, USA}
\newcommand{\myongji}{Myongji University, Yongin, Kyonggido 449-728, Korea}
\newcommand{\nagasaki}{Nagasaki Institute of Applied Science, Nagasaki-shi, Nagasaki 851-0193, Japan}
\newcommand{\newmex}{University of New Mexico, Albuquerque, New Mexico 87131, USA }
\newcommand{\nmsu}{New Mexico State University, Las Cruces, New Mexico 88003, USA}
\newcommand{\ohio}{Department of Physics and Astronomy, Ohio University, Athens, Ohio 45701, USA}
\newcommand{\ornl}{Oak Ridge National Laboratory, Oak Ridge, Tennessee 37831, USA}
\newcommand{\orsay}{IPN-Orsay, Universite Paris Sud, CNRS-IN2P3, BP1, F-91406, Orsay, France}
\newcommand{\peking}{Peking University, Beijing 100871, P.~R.~China}
\newcommand{\pnpi}{PNPI, Petersburg Nuclear Physics Institute, Gatchina, Leningrad region, 188300, Russia}
\newcommand{\riken}{RIKEN Nishina Center for Accelerator-Based Science, Wako, Saitama 351-0198, Japan}
\newcommand{\rikjrbrc}{RIKEN BNL Research Center, Brookhaven National Laboratory, Upton, New York 11973-5000, USA}
\newcommand{\rikkyo}{Physics Department, Rikkyo University, 3-34-1 Nishi-Ikebukuro, Toshima, Tokyo 171-8501, Japan}
\newcommand{\saispbstu}{Saint Petersburg State Polytechnic University, St. Petersburg, 195251 Russia}
\newcommand{\saopaulo}{Universidade de S{\~a}o Paulo, Instituto de F\'{\i}sica, Caixa Postal 66318, S{\~a}o Paulo CEP05315-970, Brazil}
\newcommand{\seoulnat}{Seoul National University, Seoul, Korea}
\newcommand{\stonybrkc}{Chemistry Department, Stony Brook University, SUNY, Stony Brook, New York 11794-3400, USA}
\newcommand{\stonycrkp}{Department of Physics and Astronomy, Stony Brook University, SUNY, Stony Brook, New York 11794-3400, USA}
\newcommand{\subatech}{SUBATECH (Ecole des Mines de Nantes, CNRS-IN2P3, Universit{\'e} de Nantes) BP 20722 - 44307, Nantes, France}
\newcommand{\tenn}{University of Tennessee, Knoxville, Tennessee 37996, USA}
\newcommand{\titech}{Department of Physics, Tokyo Institute of Technology, Oh-okayama, Meguro, Tokyo 152-8551, Japan}
\newcommand{\tsukuba}{Institute of Physics, University of Tsukuba, Tsukuba, Ibaraki 305, Japan}
\newcommand{\vandy}{Vanderbilt University, Nashville, Tennessee 37235, USA}
\newcommand{\waseda}{Waseda University, Advanced Research Institute for Science and Engineering, 17 Kikui-cho, Shinjuku-ku, Tokyo 162-0044, Japan}
\newcommand{\weizmann}{Weizmann Institute, Rehovot 76100, Israel}
\newcommand{\wigner}{Institute for Particle and Nuclear Physics, Wigner Research Centre for Physics, Hungarian Academy of Sciences (Wigner RCP, RMKI) H-1525 Budapest 114, POBox 49, Budapest, Hungary}
\newcommand{\yonsei}{Yonsei University, IPAP, Seoul 120-749, Korea}
\affiliation{\abilene}
\affiliation{\acadsin}
\affiliation{\banaras}
\affiliation{\barc}
\affiliation{\bnlcoll}
\affiliation{\bnlphys}
\affiliation{\caucr}
\affiliation{\charlesczech}
\affiliation{\chonbuk}
\affiliation{\ciae}
\affiliation{\cns}
\affiliation{\colorado}
\affiliation{\columbia}
\affiliation{\czechtech}
\affiliation{\dapnia}
\affiliation{\debrecen}
\affiliation{\elte}
\affiliation{\ewha}
\affiliation{\fit}
\affiliation{\fsu}
\affiliation{\gsu}
\affiliation{\hanyang}
\affiliation{\hiroshima}
\affiliation{\ihepprot}
\affiliation{\illuiuc}
\affiliation{\inrras}
\affiliation{\instpasczech}
\affiliation{\isu}
\affiliation{\jinrdubna}
\affiliation{\jyvaskyla}
\affiliation{\kek}
\affiliation{\korea}
\affiliation{\kurchatov}
\affiliation{\kyoto}
\affiliation{\labllr}
\affiliation{\lawllnl}
\affiliation{\losalamos}
\affiliation{\lpc}
\affiliation{\lund}
\affiliation{\maryland}
\affiliation{\mass}
\affiliation{\muenster}
\affiliation{\muhlenberg}
\affiliation{\myongji}
\affiliation{\nagasaki}
\affiliation{\newmex}
\affiliation{\nmsu}
\affiliation{\ohio}
\affiliation{\ornl}
\affiliation{\orsay}
\affiliation{\peking}
\affiliation{\pnpi}
\affiliation{\riken}
\affiliation{\rikjrbrc}
\affiliation{\rikkyo}
\affiliation{\saispbstu}
\affiliation{\saopaulo}
\affiliation{\seoulnat}
\affiliation{\stonybrkc}
\affiliation{\stonycrkp}
\affiliation{\subatech}
\affiliation{\tenn}
\affiliation{\titech}
\affiliation{\tsukuba}
\affiliation{\vandy}
\affiliation{\waseda}
\affiliation{\weizmann}
\affiliation{\wigner}
\affiliation{\yonsei}
\author{A.~Adare} \affiliation{\colorado}
\author{S.~Afanasiev} \affiliation{\jinrdubna}
\author{C.~Aidala} \affiliation{\losalamos} \affiliation{\mass}
\author{N.N.~Ajitanand} \affiliation{\stonybrkc}
\author{Y.~Akiba} \affiliation{\riken} \affiliation{\rikjrbrc}
\author{R.~Akimoto} \affiliation{\cns}
\author{H.~Al-Bataineh} \affiliation{\nmsu}
\author{H.~Al-Ta'ani} \affiliation{\nmsu}
\author{J.~Alexander} \affiliation{\stonybrkc}
\author{K.R.~Andrews} \affiliation{\abilene}
% ...
\collaboration{PHENIX Collaboration} \noaffiliation

\date{\today}


\begin{abstract}

The three $\Upsilon$ states, $\Upsilon(1S$$+$$2S$$+$$3S)$, are measured in 
$d$+Au and $p$$+$$p$ collisions at $\sqrt{s_{_{NN}}}$=200 GeV and 
rapidities $1.2 < |y| < 2.2$ by the PHENIX experiment at the Relativistic 
Heavy-Ion Collider.  Cross sections for the inclusive 
$\Upsilon(1S$$+$$2S$$+$$3S)$ production are obtained.  The inclusive yields 
per binary collision for $d$+Au collisions relative to those in $p$$+$$p$ 
collisions ($R_{dAu}$) are found to be 0.62 $\pm$ 0.26 (stat) $\pm$ 0.13 
(syst) in the gold-going direction and 0.91 $\pm$ 0.33 (stat) $\pm$ 0.16 
(syst) in the deuteron-going direction.  The measured results are compared 
to a nuclear-shadowing model, EPS09 [JHEP {\bf 04}, 065 (2009)], combined 
with a final-state breakup cross section, $\sigma_{br}$, and compared to 
lower energy $p$+A results.  We also compare the results to the PHENIX 
$J/\psi$ results [Phys. Rev. Lett. {\bf 107}, 142301 (2011)].  The rapidity 
dependence of the observed $\Upsilon$ suppression is consistent with 
lower energy $p$+A measurements.

\end{abstract}

	


\maketitle

\section{Introduction \label{sec:intro}} 

Quarkonia are produced dominantly by the gluon-gluon fusion process in high 
energy collisions~\cite{productionmechanism2,productionmechanism3}. 
Therefore, quarkonia production is a good probe to explore the gluon 
distribution of the nucleon and its modification in nuclei. Recently, the 
PHENIX collaboration has reported $J/\psi$ suppression in \sqsn=200 GeV 
deuteron-gold (\dau) collisions at the Relativistic Heavy Ion 
Collider~\cite{jpsirdau2011}. The centrality dependence of these $J/\psi$ 
suppression results at forward rapidity is not well described quantitatively 
by nuclear-shadowing models that include final-state breakup 
effects~\cite{jpsimodel}. Because the \ups mass is heavier than 
$J/\psi$, the nuclear effects on the gluon distribution can be studied in 
different kinematic regions.  At forward rapidity (the deuteron going direction) 
and the same collision energy of \sqsn=200 GeV, the average momentum fraction 
of the gluon in the gold nucleus that is sampled for \ups production is 
$\left<x_2\right>$ $\approx$ 1 $\times$ $10^{-2}$, 
whereas $J/\psi$ production samples 
$\left<x_2\right>$ $\approx$ 3 $\times$ $10^{-3}$.

There are various fits for the nuclear parton distribution functions (nPDFs) 
over broad $x$ ranges~\cite{eps09,eks98_1,eks98_2,ekps,nds}. In \dau 
collisions, since the forward and backward rapidities cover different $x_2$ 
($x$ in the Au nucleus) ranges, \ups production at these two rapidities 
would be affected differently by these nPDFs. Additionally, the final-state 
breakup effect should also suppress \ups yields by some amount at both 
rapidities, but there is no clear indication of the size of this effect 
yet~\cite{ramonardau}. Thus, \ups measurements in \dau collisions should 
give new and valuable information to test nuclear parton modification and 
breakup effects.

Lattice quantum chromodynamics predicts that the $\uone$, $\utwo$, and 
$\uthree$ all have different binding energies and radii, and so should melt 
at different temperatures of the hot nuclear medium~\cite{helmutsatz}. 
Therefore, the three \ups states are thought to be good probes for the 
temperature of the hot dense matter. Recently, the Compact Muon Solenoid 
(CMS) experiment at the Large Hadron Collider (LHC) reported that the double 
ratio of the \utt excited states to the $\uone$ ground state in \pbpb and 
\pp collisions at \sqsn=2.76 TeV,

\begin{eqnarray}
{\frac{\utt/\uone|_{\pbpb}}{\utt/\uone|_{\pp}}} = 
0.31^{+0.19}_{-0.15}\rm{(stat)} \pm 0.03\rm{(syst)} \nonumber
\end{eqnarray}

\noindent for single decay muons of $p_T$ $>$ 4 GeV/$c$ and $|\eta|$ $<$ 
2.4~\cite{CMSupsilon}. They also reported $\uone$ is suppressed by 
approximately 40$\%$ in minimum-bias \pbpb collisions~\cite{CMSupsilon}.

In \ups suppression for nucleus-nucleus collisions, there should be 
contributions from cold nuclear matter as well as those from the hot nuclear 
matter. Thus, to separate these two types of contributions, it is necessary 
to measure the level of suppression from cold nuclear matter effects with 
$p(d)$+A collisions, where hot nuclear matter is not created.

A lower-energy fixed-target experiment, E772, reported measurements in 
\sqsn=38.8 GeV $p$ + A collisions of the $\uone$ ground state and the $\utt$ 
excited states. The observed suppression of the $\uone$ and $\utt$ agree 
within the experimental uncertainties~\cite{E772}. The initial-state effects 
from nuclear shadowing are not expected to differ between the three \ups 
states since they are produced mostly by gluon-gluon fusion subprocesses and 
have similar 
masses~\cite{productionmechanism2,productionmechanism3,ramonardau}. For the 
final-state breakup effect, there is no clear estimate of its energy 
dependence and of the difference between the three \ups states. In this 
paper, we present the first measurement of inclusive $\uall$ cold nuclear 
matter effects as well as the production cross section using \dau and \pp 
collisions at \sqsn=200 GeV measured by the PHENIX experiment.

\section{Analysis Method  \label{sec:analysis}} 

\subsection{Experimental setup \label{sec:exp}}

\begin{figure}[]
  \includegraphics[width=1.0\linewidth]{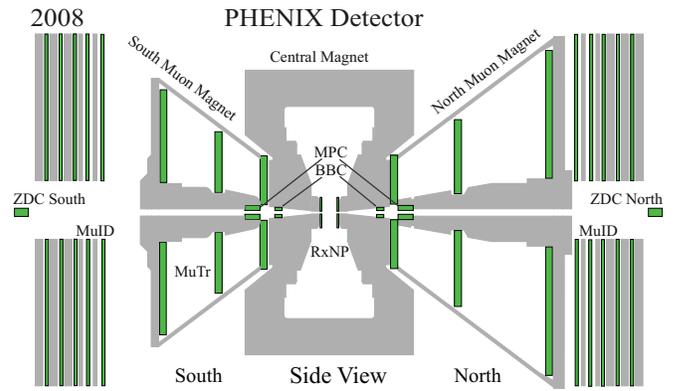}
  \caption{\label{fig:phenix2008} (color online) PHENIX detector
  configuration in 2008. This side view includes the forward-rapidity detectors
  (South and North arms): the Muon Tracker (MuTr), the Muon Identifier (MuID) 
  for muon detection and identification, and the beam-beam counter (BBC) 
  for global event characteristics.}  
\end{figure}

The PHENIX apparatus is described in detail in Ref.~\cite{phenixdetector}. 
In \dau collisions, the deuteron comes from the negative-rapidity end of 
PHENIX (South) and goes towards positive-rapidity (North), and vice versa 
for the gold ions. For the \ups analysis presented here, three detector 
systems are required for reconstruction and triggering at forward and 
backward rapidities. These are the Muon Tracker (MuTr), the Muon Identifier 
(MuID), and the beam-beam counters (BBCs). There are two separate BBC 
systems. One covers forward rapidity and the other covers backward rapidity 
as shown in Fig.~\ref{fig:phenix2008}.

Each BBC comprises 64 quartz $\breve{\rm C}$erenkov radiators and mesh 
dynode PMTs. The two BBCs are located at $\pm$144~cm from the nominal 
interaction point and cover pseudorapidity of $3 < |\eta| < 3.9$. Each MuID 
comprises five layers of thick steel vertical plates with Iarocci tubes 
between each pair of plates. Most hadrons are absorbed in the steel plates. 
Muons with more than 2.7 GeV/$c$ of momentum will pass through all layers of 
the MuID and reach the last gap. Each MuTr is composed of three stations of 
cathode strip chambers and measures the momentum and charge sign of the muon 
according to their bending in the magnetic field, with coverage in rapidity 
of $1.2 < |y| < 2.2$ for the \ups and full azimuthal coverage of $\phi$ 
$\in$ $[-\pi,\pi]$. The nose-cone absorber and the central-magnet pole face, 
which both lie between the interaction region and the innermost part of the 
muon tracker, also help to reduce hadron backgrounds, especially by 
eliminating many light hadrons (e.g. $\pi$, $K$) before they decay into 
secondary muons. Fewer than 1\% of hadrons punch through the absorbers, 
reach the last gap of MuID, and become fake muon tracks.

The data sets used in this analysis were collected during 2006, 2008, and 
2009 using the BBC Level-1 trigger. This BBC trigger requires hits in the 
negative and positive rapidity ends of the BBC in order to register an 
interaction and provide a minimum-bias trigger. The BBC also measures the 
z-vertex position of the interaction using time differences between its hits 
in the negative and positive rapidity directions. For this analysis, the 
z-vertex is required to be within $\pm$30~cm of the center of PHENIX, z~=~0. 
Additionally, the MuID Level-1 trigger is used in order to require that at 
least two particles penetrate through the MuID to its last layer.

After removing bad runs, such as those with numerous high-voltage trips and 
significant detector performance variations, the integrated luminosities of 
the collected data are 69~nb$^{-1}$ and 67~nb$^{-1}$ for the positive and 
negative rapidity muon detectors in \dau collisions from 2008. Here, 
69~nb$^{-1}$ and 67~nb$^{-1}$ correspond to 27.2~pb$^{-1}$ and 
26.4~pb$^{-1}$ when scaled by the number of participants. For the \pp 
collisions, the integrated luminosities are 22.5 pb$^{-1}$ and 22.2 
pb$^{-1}$ for the positive and negative rapidities from 2006 and 2009.

We apply quality-assurance cuts on the data to select good tracks and 
improve the signal-to-background ratio. We calculate the track $\chi^2$ and 
vertex $\chi_{\rm vtx}^{2}$, and match the tracks in the MuID and the MuTr 
at the first layer of the MuID in both position and angle. We also check the 
number of hits in a MuID road, which is a straight line that connects sets 
of hits in different layers of the MuID. We compare momenta of the two muons 
and remove pairs with a large asymmetry ($|(p_1-p_2)/(p_1+p_2)| > 0.6$) 
between the two momenta. These asymmetric-momenta pairs are largely from 
random pairs where one hadron has decayed into a muon inside the tracking 
volume and has been misreconstructed as a higher momentum track; thus 
yielding a fake high-mass pair. The efficiency loss from this cut for \upss 
is less than 2\%. The values of the cuts are determined using the PHENIX 
{\sc geant}3-based~\cite{geant} ({\sc PISA}) detector simulations.

\begin{figure*}[!h]
  \includegraphics[width=0.49\linewidth]{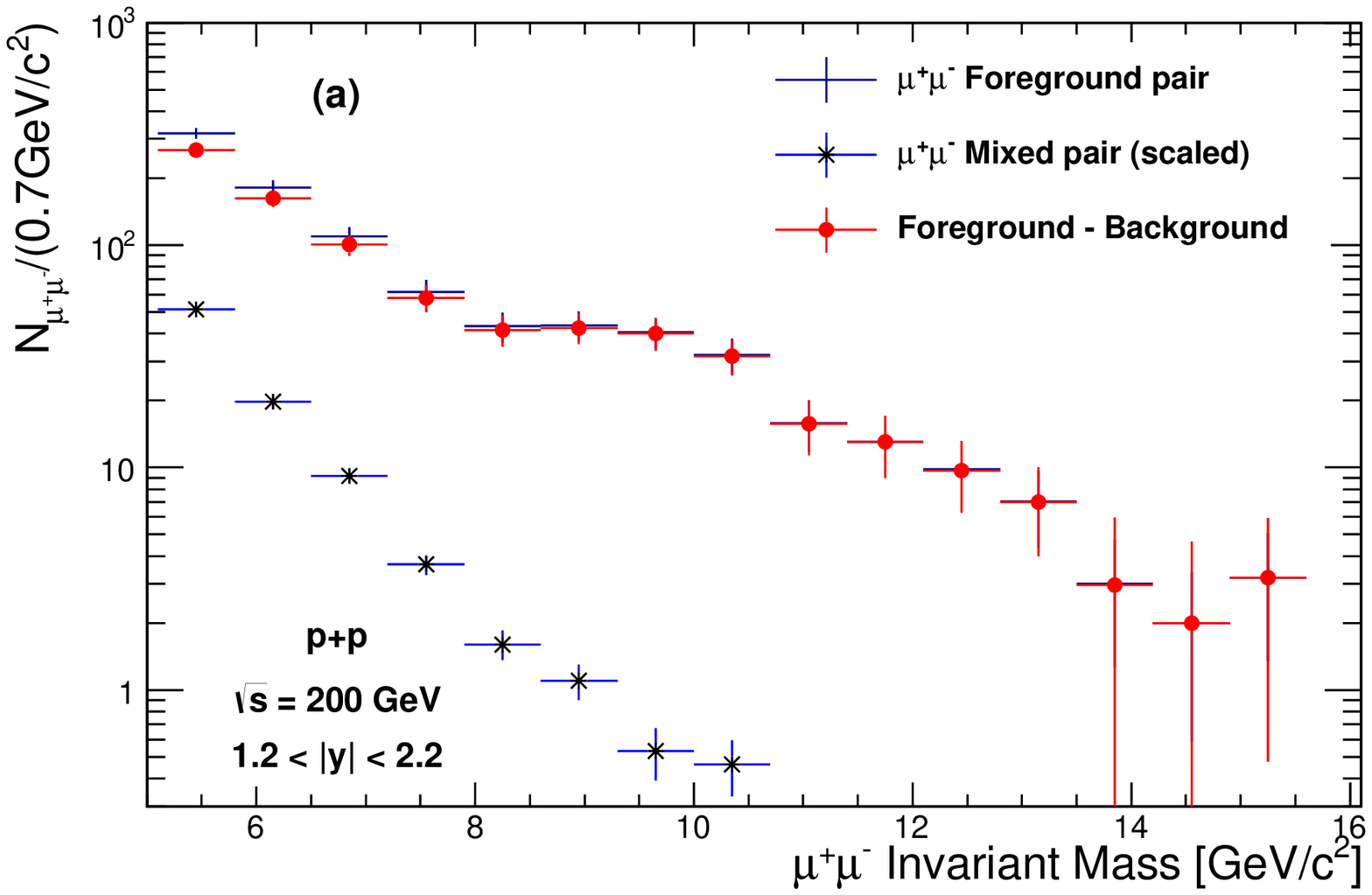}
  \includegraphics[width=0.49\linewidth]{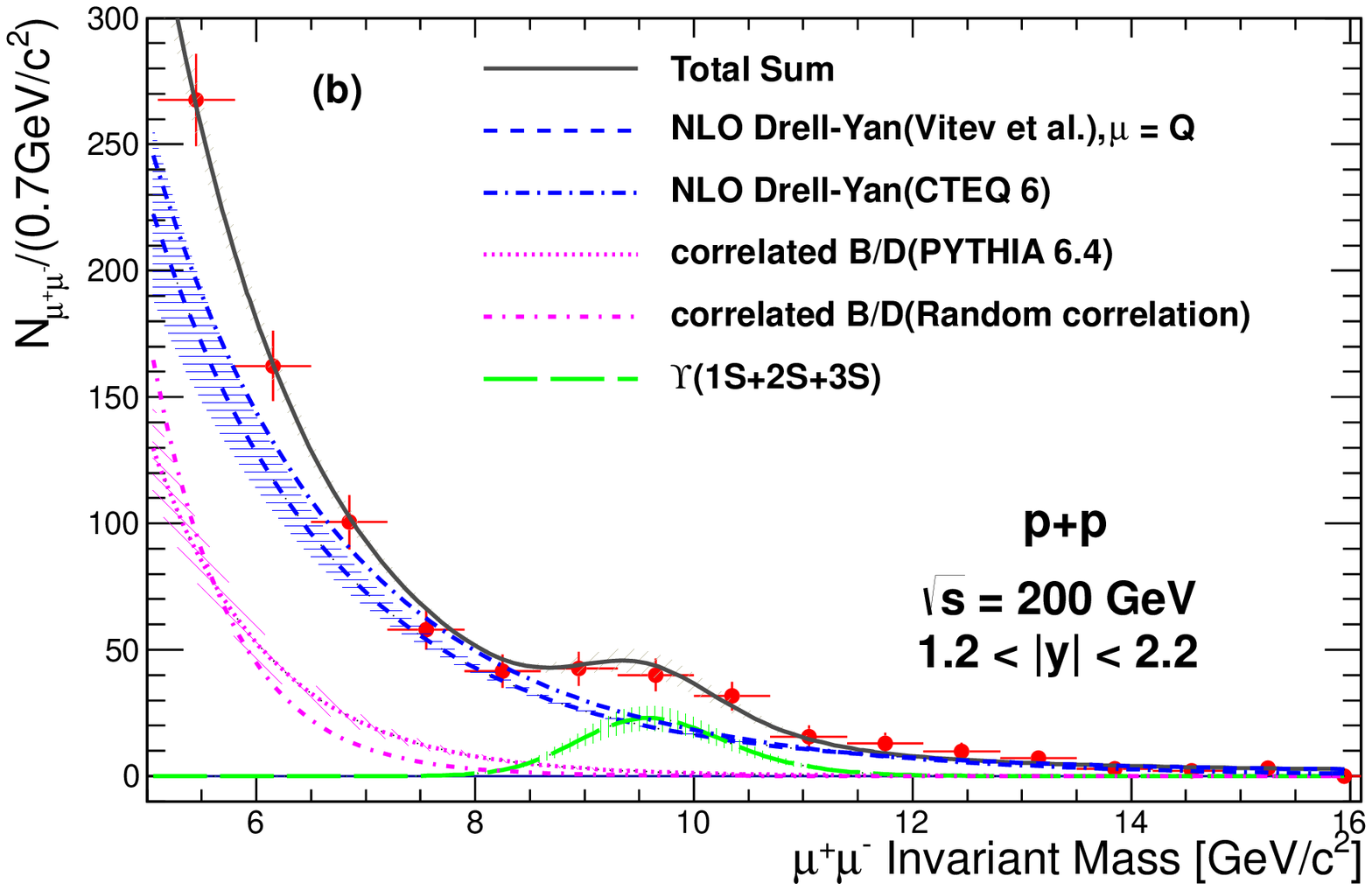}
  \includegraphics[width=0.49\linewidth]{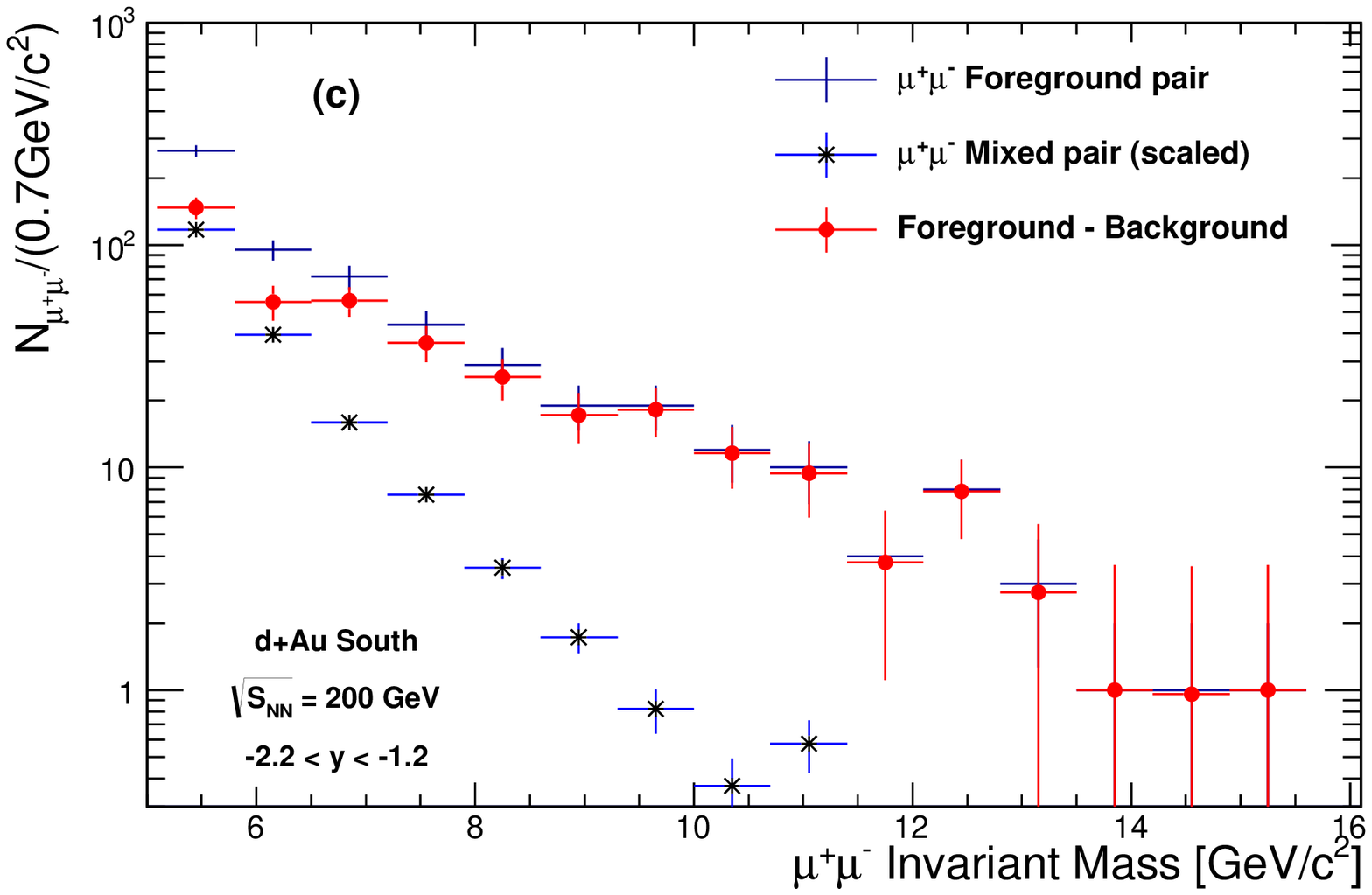}
  \includegraphics[width=0.49\linewidth]{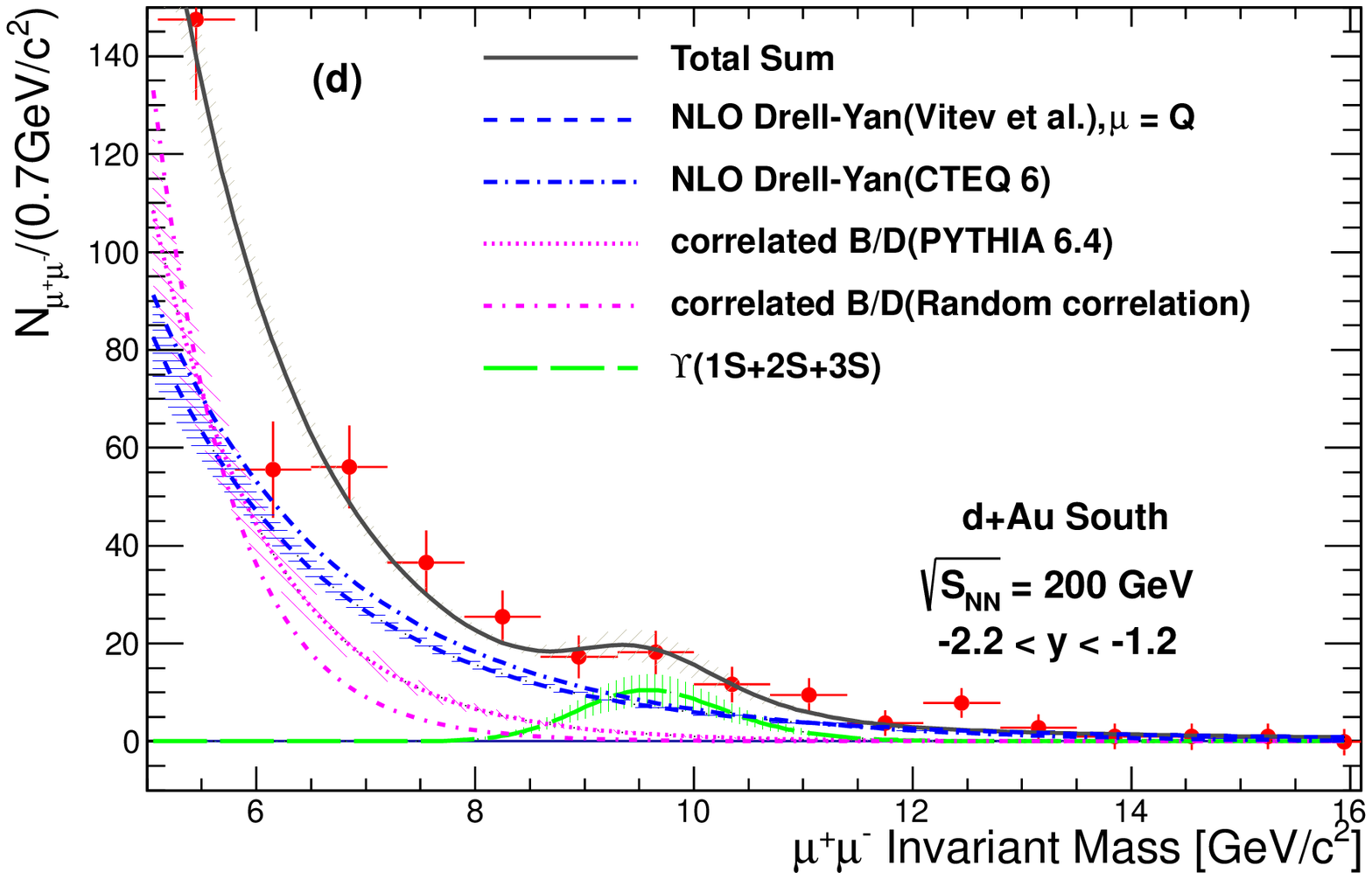}
  \includegraphics[width=0.49\linewidth]{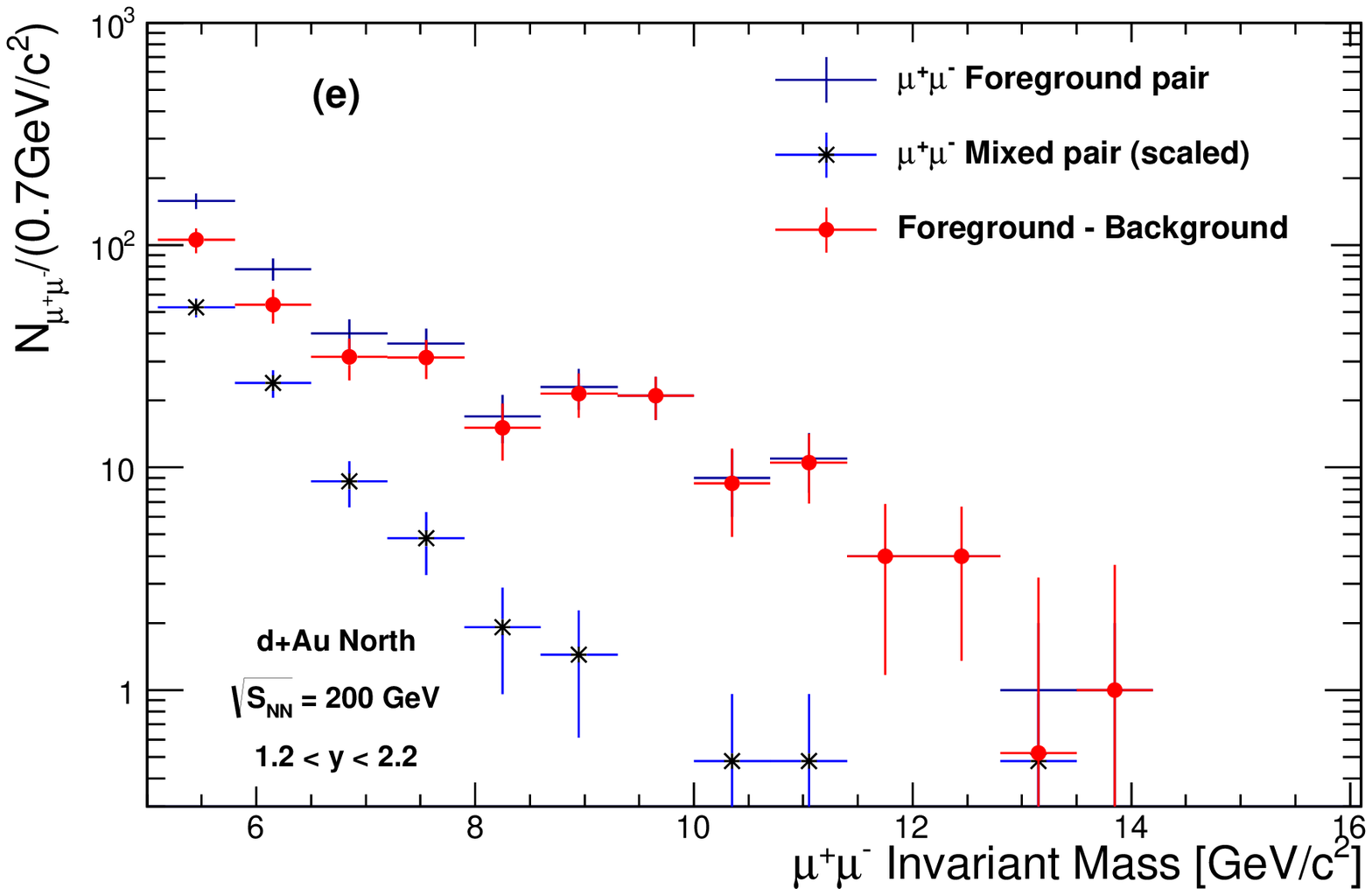}
  \includegraphics[width=0.49\linewidth]{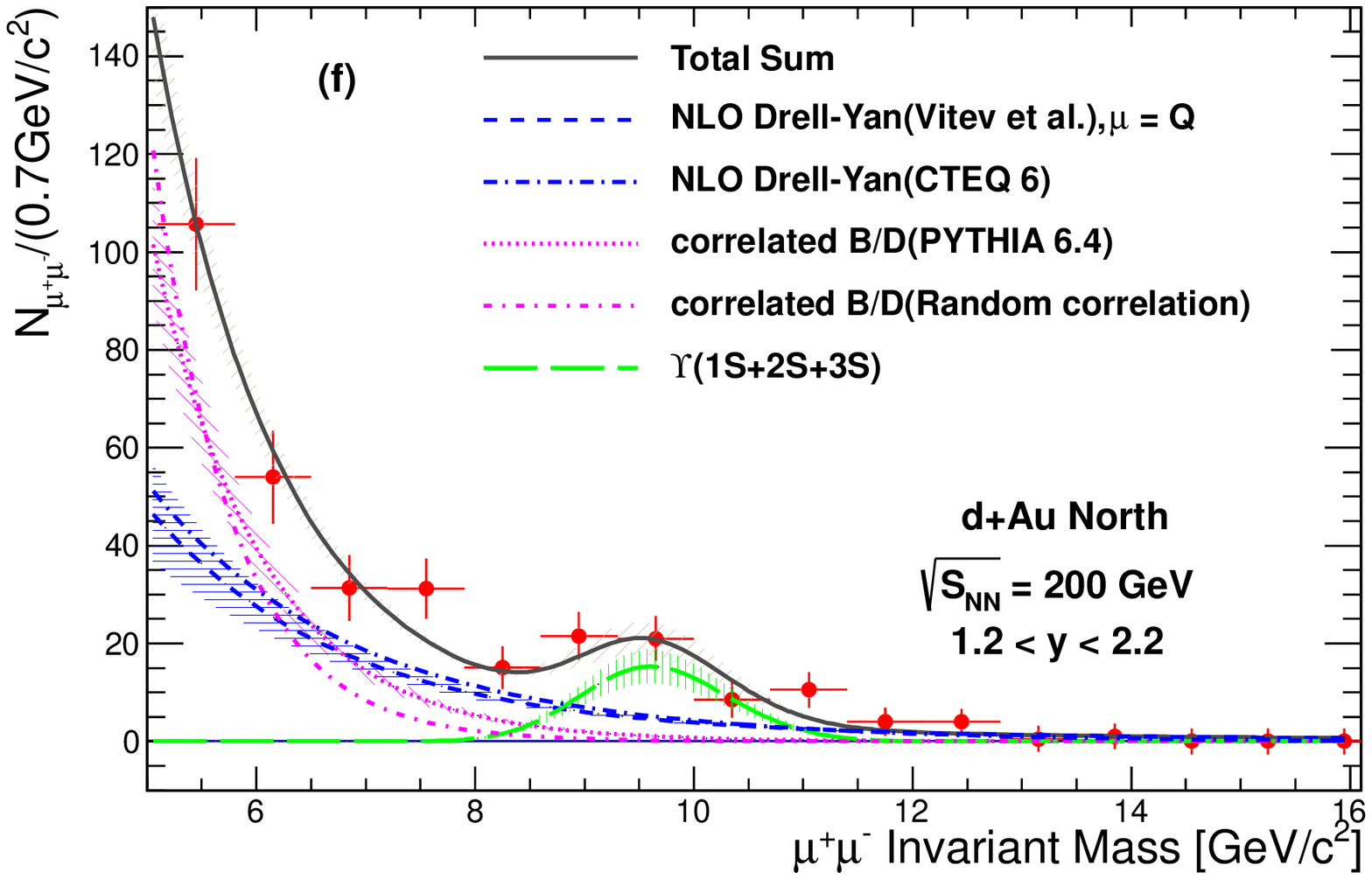}
  \caption{\label{fig:fgbg} (color online) The invariant mass distributions
  between 5 GeV/$c^{2}$ and 16 GeV/$c^{2}$ are shown for \pp collisions (a,b) 
  and in \dau collisions for the South arm (c,d) and the North arm (e,f). 
  In the mass distributions, (a), (c), and (e) show the mass distribution 
  of unlike-sign foreground pairs, the mixed event pairs as combinatorial 
  background, and the subtraction of background pairs from foreground pairs.
  (b), (d), and (f) show the combinatorial background subtracted signal 
  overlaid with the correlated backgrounds and $\uall$.   
  The shaded bands around the curves represent the uncertainties
  from the fitting or calculations of the renormalization and factorization 
  scales, Q/2 $\leq \mu \leq$ 2Q.
  }
\end{figure*}

For this analysis, we form an invariant-mass distribution from the 
unlike-charge-sign (foreground) pairs of muon tracks. In addition to the 
quarkonia resonances including the \ups signal, the mass distribution also 
contains uncorrelated (combinatorial) background and correlated background 
pairs. There are two methods to estimate the combinatorial backgrounds: 1) 
use like-sign pairs of muons from the same event, or 2) use an event-mixing 
method which mixes unlike-sign muons from different events to form random 
pairs. In this analysis, we use the event-mixing method to estimate the 
combinatorial background as shown in Fig.~\ref{fig:fgbg} (a), (c) and (e), 
and assign a systematic uncertainty based on the difference between the two 
methods.  We calculate the normalization factor for the mixed events by
\begin{equation}
{\rm Normalization}\,\,{\rm Factor} = 
\frac{2 \times \sqrt{FG_{++} \times FG_{--}}}{BG_{+-}}. 
\label{eq:mixedeventnormalization}
\end{equation}
Here, $BG_{+-}$ stands for the number of the unlike-sign mixed events, 
$FG_{++}$ and $FG_{--}$ represent the number of the like-sign events. 
Unlike-sign mixed events are scaled by the normalization factor and we 
assign a 3\% systematic uncertainty for this factor.
   
After the combinatorial background is subtracted, in the \ups-mass region, 
there are still contributions from correlated backgrounds expected from the 
Drell-Yan process and pairs of muons from the same $c\overline{c}$ or 
$b\overline{b}$ pairs. Therefore, it is important to estimate the correlated 
backgrounds properly to extract the \ups signal. We use next-to-leading 
order (NLO) calculations and {\sc pythia} 6.4~\cite{pythia64} to estimate 
these correlated backgrounds, and the PHENIX {\sc geant}3 simulation to 
include a realistic detector response as shown in Fig.~\ref{fig:fgbg} (b), 
(d) and (f). Details of the estimates for these correlated backgrounds are 
described in the following sections.\\

\subsection{\ups and physical background estimation \label{sec:simulation}}

\subsubsection{The Drell-Yan process \label{sec:drellyan}}

\begin{figure}[!h]
  \includegraphics[width=1.0\linewidth]{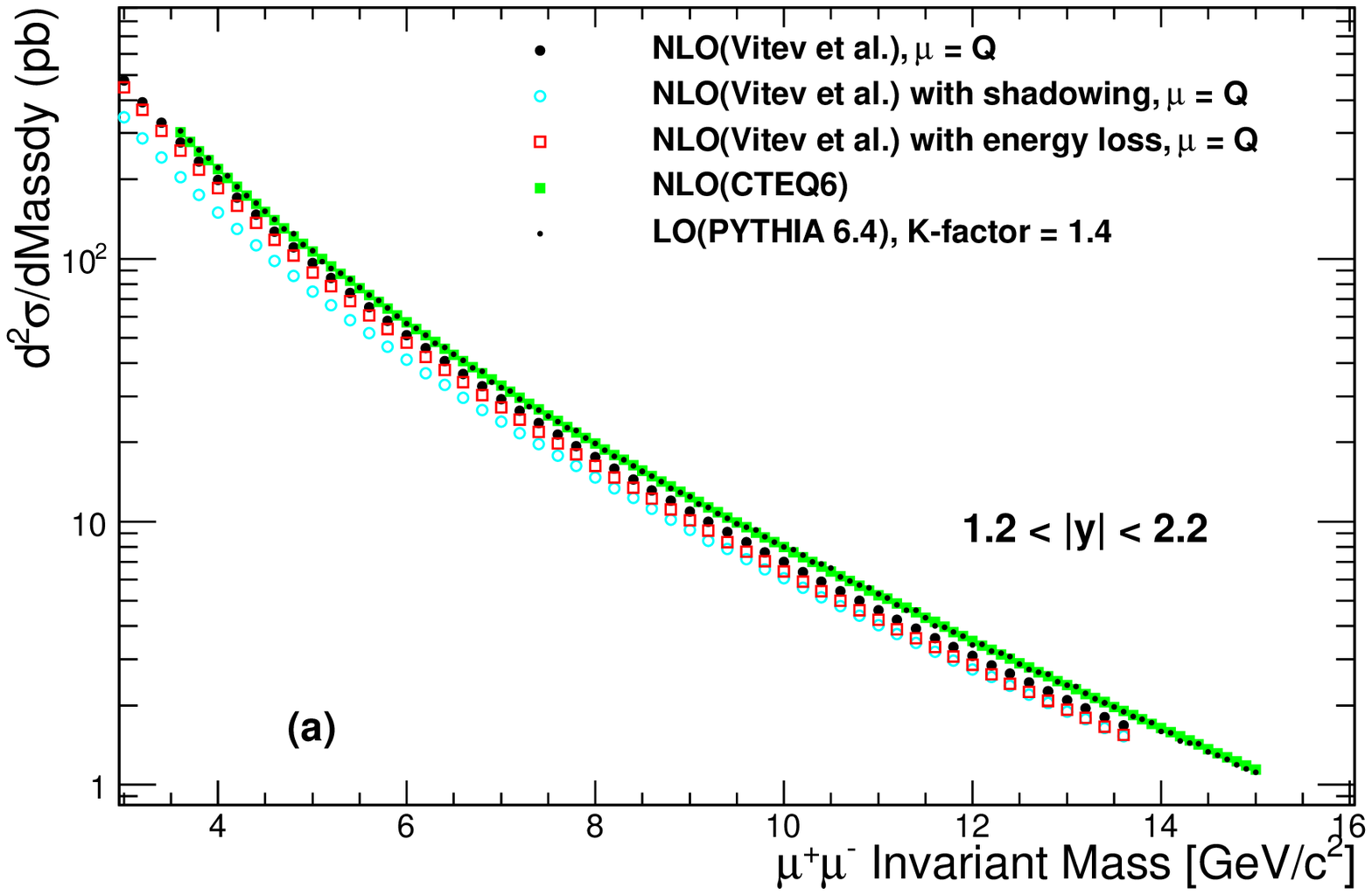}
  \includegraphics[width=1.0\linewidth]{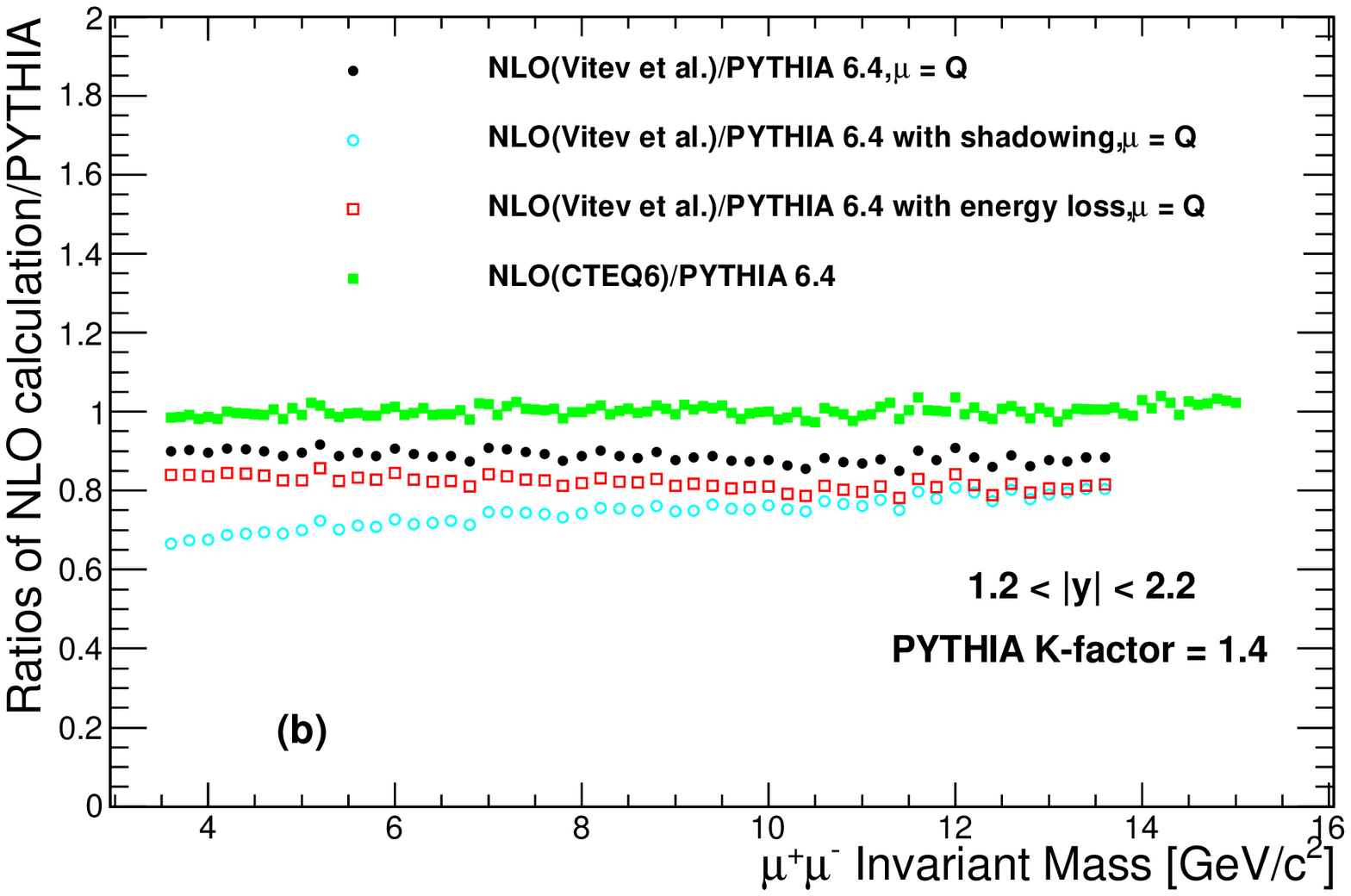}
  \caption{\label{fig:dyrat} (color online) Differential cross sections for the 
  Drell-Yan process,
  $q\overline{q}$ $\rightarrow$ $\gamma^{*}$ $\rightarrow$ $\mu^{+}\mu^{-}$
  are drawn (a) for a {\sc pythia} calculation and for NLO 
  calculations~\cite{Ivan,cteqnlo} 
  in the rapidity region $1.2 < |y| < 2.2$.  The ratios of the NLO 
  calculations over that from {\sc pythia} 6.4 ($K$-factor=1.4) 
  are also shown (b).}
\end{figure}

The mass region between 4 and 8 GeV/$c^{2}$ (above the $J/\psi$, $\psi^{'}$ 
masses and below the \ups mass) is dominated by the Drell-Yan process and by 
correlated open-heavy flavor pairs.  The very low statistics above the \ups 
mass, where the Drell-Yan process dominates, does not provide a useful 
constraint on the Drell-Yan yield; so, we use NLO calculations from 
Vitev~\cite{Ivan} to constrain the Drell-Yan yields and to estimate their 
contribution in the \ups-mass region. NLO calculations of the Drell-Yan 
process are known to be very accurate from comparisons to data at other 
energies~\cite{Ivan,fnaldrellyan,cdfdrellyan}. For \dau collisions, nuclear 
effects are added in the NLO calculations - including isospin effects which 
account for the composition of the nucleus in terms of neutrons and protons, 
parton shadowing corrections, and the effect of initial-state energy 
loss~\cite{Ivan}.

To evaluate the model's systematic uncertainty for the Drell-Yan 
contribution, we use a calculation from CTEQ~\cite{cteqnlo}, as shown in 
Fig.~\ref{fig:dyrat}a.  For the Vitev calculation~\cite{Ivan} without 
nuclear corrections, the difference between the CTEQ calculation and that 
from Vitev is approximately 10\% over the entire mass range (filled-green 
squares and the black circles in Fig.~\ref{fig:dyrat}b).  We assume this 
same systematic uncertainty for \pp and \dau collisions.  Additionally, the 
variation of the renormalization and factorization scales in the calculation 
from Vitev, Q/2 $\leq \mu \leq$ 2Q, is included as a systematic uncertainty 
for both collision systems.

The Drell-Yan contribution to the data is determined using the calculated 
cross section and the integrated luminosity for each data set.  This 
contribution is corrected for geometrical acceptance and efficiencies. 
Details are shown in 
Eq.~(\ref{eq:dycalculation}, \ref{eq:dycalculation2}, 
and (\ref{eq:dycalculation3}); and are discussed in the text that follows.

\begin{equation}
\frac{d\sigma_{\rm DY}}{dm} \cdot \mathcal{L} \cdot \epsilon^{\rm BBC}_{\rm DY} \cdot A \epsilon_{\rm DY} = \frac{N_{\rm DY}}{\Delta m},
\label{eq:dycalculation}
\end{equation}
\begin{equation}
\frac{d\sigma_{\rm DY}}{dm} \cdot \frac{N_{\rm MB}}{\sigma_{\rm Tot} \cdot \epsilon^{\rm BBC}_{\rm MB}} \cdot \epsilon^{\rm BBC}_{\rm DY} \cdot A \epsilon_{\rm DY} = \frac{N_{\rm DY}}{\Delta m},
\label{eq:dycalculation2}
\end{equation}
\begin{equation}
\frac{d\sigma_{\rm DY}}{dm} \cdot \frac{N_{\rm MB}}{\sigma_{\rm Tot} \cdot C} \cdot A \epsilon_{\rm DY} = \frac{N_{\rm DY}}{\Delta m},
\label{eq:dycalculation3}
\end{equation}

\noindent

where ${d\sigma_{\rm DY}}/{dm}$ is the differential cross section of the 
Drell-Yan process, $q\overline{q}$ $\rightarrow$ $\gamma^{*}$ $\rightarrow$ 
$\mu^{+}\mu^{-}$, from the NLO calculation for each mass bin in the rapidity 
region $1.2 <|y|< 2.2$. $\mathcal{L}$ stands for the integrated 
luminosity, $N_{\rm MB}/(\sigma_{\rm Tot} \cdot \epsilon^{\rm BBC}_{\rm 
MB})$, where $N_{\rm MB}$ stands for the number of sampled minimum-bias (MB) 
events and $\sigma_{\rm Tot}$ represents the total inelastic BBC MB cross 
section, 42.2~mb (2260~mb) for \pp (\dau) collisions. $\epsilon^{\rm 
BBC}_{\rm MB}$ and $\epsilon^{\rm BBC}_{\rm DY}$ are the BBC trigger 
efficiencies for MB events and Drell-Yan events, respectively. 
$C=\epsilon_{\rm MB}^{\rm BBC}$/$\epsilon_{\rm DY}^{\rm BBC}$ is a 
correction factor for the relative BBC efficiencies of minimum bias compared 
to hard processes containing a Drell-Yan pair. Its value is determined using 
a Glauber model and a simulation of the BBC, and is 0.69 (0.89) for \pp 
(\dau) collisions. $A\epsilon_{\rm DY}$ represents the product of the 
detector acceptance and efficiency, including the effect of the Level-1 
trigger. Finally, ${N_{\rm DY}}/{\Delta m}$ is the yield of dimuon pairs 
from the Drell-Yan process for each mass bin.

The detailed procedure to estimate the Drell-Yan yields for \pp and \dau, 
using Eq.~(\ref{eq:dycalculation}) is as follows. First, we generate the 
correct number of Drell-Yan events, which we estimate by multiplying the 
differential cross section by the accumulated luminosity for each invariant 
mass bin considering BBC efficiencies. This corresponds to ${d\sigma_{\rm 
DY}}/{dm} \cdot \mathcal{L} \cdot \epsilon^{\rm BBC}_{\rm DY}$ in Eq. 
(\ref{eq:dycalculation}). For example, $\mathcal{L}$ is 22.5 pb$^{-1}$ for 
the forward-rapidity \pp data. After event generation, to account for the 
acceptance times efficiency, $A \epsilon_{\rm DY}$, the generated Drell-Yan 
events from the luminosity-weighted NLO calculation are then run through the 
PHENIX {\sc geant}3 simulation and are reconstructed in the same way as real 
data. In the simulation, hit positions in all the muon detectors are 
registered and are reconstructed, including the effects of disabled HV 
channels and detector efficiencies. The resulting simulated counts in mass 
bins, ${N_{\rm DY}}/{\Delta m}$, are then fit with an exponential function. 
This function describes the simulated distribution very well with a fit 
quality of $\chi^2$ per degree-of-freedom ($\chi^2$/dof) of 34.9/36 to 
38.9/36. The shape and yield of this function are then fixed and used in the 
fits to the data, and represent the contribution of Drell-Yan in the fit 
function, Eq. (\ref{eq:fittingforupsilon}).

Although the NLO cross sections for \dau collisions already include nuclear 
corrections, they are still per nucleon-nucleon collision, and need to be 
scaled up by the number of binary collisions, $N_{\rm coll}$. Eq. 
(\ref{eq:dydauscale}) shows the relation between the cross section for \pp 
and that for \dau collisions, which is derived from Eq. 
(\ref{eq:upsilonrdau}).

\begin{equation}
R_{d\rm Au} = 1 = \frac{dN_{\rm DY}^{d\rm Au}/dm}{\left<N_{\rm coll}\right>dN_{\rm DY}^{pp}/dm}, \nonumber \\
\label{eq:upsilonrdau}
\end{equation}
\begin{equation}\frac{\sigma_{\rm Tot}^{d\rm Au}}{\sigma_{\rm Tot}^{pp}} = \frac{(dN_{\rm DY}^{d\rm Au}/dm) \cdot \sigma_{\rm Tot}^{d\rm Au}}{\left<N_{\rm coll}\right>(dN_{\rm DY}^{pp}/dm) \cdot \sigma_{\rm Tot}^{pp}} = \frac{d\sigma_{\rm DY}^{d\rm Au}/dm}{\left<N_{\rm coll}\right>d\sigma_{\rm DY}^{pp}/dm}, \nonumber \\
\label{eq:crosssectionexpansion}
\end{equation}
\begin{equation}
d\sigma_{\rm DY}^{d\rm Au}/dm = \left<N_{\rm coll}\right>d\sigma_{\rm DY}^{pp}/dm\frac{\sigma_{\rm Tot}^{d\rm Au}}{\sigma_{\rm Tot}^{pp}},
\label{eq:dydauscale}
\end{equation}

\noindent
where $dN_{\rm DY}^{d\rm Au (pp)}/dm$ is the invariant yield for the 
Drell-Yan process in \dau (\pp) collisions and $\sigma_{\rm Tot}^{d\rm Au 
(pp)}$ is the total inelastic cross section for \dau (\pp) collisions. In 
the expansion of Eq. (\ref{eq:crosssectionexpansion}), $dN_{\rm DY}^{d\rm Au 
(pp)}/dm \cdot \sigma_{\rm Tot}^{d\rm Au (pp)} $ is the differential cross 
section for the Drell-Yan process, $d\sigma_{\rm DY}^{d\rm Au (pp)}/dm$.  
$\left<N_{\rm coll}\right>$ is the mean number of binary collisions and is 
calculated using a Glauber model and a simulation of the BBC. $\left<N_{\rm 
coll}\right>$ is 7.6 $\pm$ 0.4 for inclusive $d$~+~Au collisions. Eq. 
(\ref{eq:dydauscale}) is used for the Drell-Yan estimates in \dau collisions 
and $d\sigma_{\rm DY}^{pp}/dm$ is considered as nuclear-effect-corrected 
cross sections per nucleon-nucleon before scaling up.

\subsubsection{Correlations of open heavy-flavor pairs \label{sec:opencorrelation}}

\begin{table}[t]
\caption{Simulation parameter settings for open beauty(charm) production.
We used {\sc pythia} 6.4 with the CTEQ5L parton distribution 
functions~\cite{cteq5l}. The {\sc pythia} tunes are from a PHENIX dilepton 
mass spectra study~\cite{Adare:dilepton}}
\label{table:open_simulation}
\begin{ruledtabular}
\begin{tabular*}{\linewidth}{@{\extracolsep{\fill}}cc}
Name of parameter & Setting\\
\hline
Bottom (Charm) Quark production & on\\
Bottom (Charm) Quark mass & 4.1 (1.25) GeV/$c^{2}$ \\
$k_{T}$ & 1.5 GeV/$c$\\
$K$-factor & 3.4 \\
$Q^2$ & 4 GeV$^{2}$ \\
\end{tabular*}
\end{ruledtabular}
\end{table}

Several measurements of open bottom and charm cross sections have been made 
by PHENIX. A recent single-electron measurement of heavy-quark production at 
midrapidity obtained $\sigma_{c\overline{c}}$=551 $\pm$~57~(stat) 
$\pm$~195~(syst)~$\mu$b~\cite{Adare:electronHF} for the total charm cross 
section. A dielectron measurement of the continuum charm pairs showed a 
total cross section of $\sigma_{c\overline{c}}$=544 $\pm$~39~(stat) 
$\pm$~142~(syst) $\pm$~200~(model)~$\mu$b~\cite{Adare:dilepton}. A 
perturbative-quantum-chromodynamics fixed-order-next-to-leading-log 
calculation~\cite{Adare:fonll} predicts a cross section of 256 
$^{+400}_{-146}$~$\mu$b, which is within experimental and theoretical 
uncertainties with these measurements.

Existing bottom cross section measurements from PHENIX also agree within 
their uncertainties. An electron-hadron charge correlation measurement 
showed a total bottom cross section of $\sigma_{b\overline{b}}$=3.2 
$^{+1.2}_{-1.1}$~(stat) 
$^{+1.4}_{-1.3}$~(syst)~$\mu$b~\cite{Adare:electronhadroncorrelation} and a 
continuum mass distribution study obtained $\sigma_{b\overline{b}}$=3.9 
$\pm$~2.5~(stat) $^{+3}_{-2}$~(syst)~$\mu$b~\cite{Adare:dilepton}. Meanwhile 
the calculation~\cite{Adare:fonll} predicts $\sigma_{b\overline{b}}$=1.87 
$^{+0.99}_{-0.67}$~$\mu$b~\cite{Adare:fonll}, consistent with the 
measurements.

To obtain an estimate of the mass shape and to generate simulated charm- and 
bottom-pair background events (see Table~\ref{table:open_simulation} for the 
simulation settings), we use the {\sc pythia} 6.4 tune, the same as that 
used for the PHENIX dilepton mass spectrum study~\cite{Adare:dilepton}. The 
generated events are run through the PHENIX {\sc geant}3 simulation to 
account for the detector acceptance at forward-rapidity and are 
reconstructed by using identical code to that used to reconstruct the data, 
with the detector efficiencies included. Before reconstructing the simulated 
events, they are embedded into real events in order to match and evaluate 
the effects of the multiplicity that exists for the data.

The resulting mass spectrum is fit by an exponential function with 
($\chi^2$/dof) of 10.0/12 to 11.8/12, which is then used to represent the 
open heavy-flavor component in the fits to the data, as in Eq. 
(\ref{eq:fittingforupsilon}). The shape is fixed and the normalization, or 
yield, is allowed to vary in the fits to the data. The data fit-values 
obtained are within a half sigma in the experimental uncertainties of the 
previously measured charm and bottom cross sections~\cite{Adare:dilepton} 
(see Section~\ref{sec:fitting} for further detail on the fits). For the 
shape of the fit function, we assign a systematic uncertainty by varying the 
slopes by $\pm$10$\%$ from the nominal values obtained from the simulation.

The relative ratio of bottom and charm production is fixed according to the 
measured production cross sections obtained in the PHENIX dilepton mass 
spectrum study~\cite{Adare:dilepton} with $\sigma_{c\overline{c}}$=544 $\pm$ 
39 (stat) $\pm$ 142 (syst) $\pm$ 200 (model) $\mu$b and 
$\sigma_{b\overline{b}}$=3.9 $\pm$ 2.5 (stat) $^{+3}_{-2}$ (syst) $\mu$b. 
Since these have large measurement uncertainties, we assign a systematic 
uncertainty for the relative ratio of bottom and charm production cross 
sections by varying this ratio by $\pm$100$\%$ from the nominal value; 
however, this does not result in a significant difference for the \ups yield 
because the charm contribution is negligible in the \ups-mass region.

Finally, a random angular correlation of the two open bottoms that form a 
pair is considered as an extreme case for the bottom correlation since NLO 
effects or interactions with other particles could alter the muon directions 
and destroy the angular correlation of the two heavy quarks. The $p_{T}$ 
spectra of the single muons from open heavy-flavor decay are sampled and the 
azimuthal correlation angle $\phi$ of the decay muons is randomized for each 
muon and then pairs are formed, effectively destroying the angular 
correlation. The resulting difference between {\sc pythia} estimation and 
random correlation is assigned as an additional systematic uncertainty.

\subsubsection{\ups estimation \label{sec:upsilon}}

\begin{figure}[!h]
  \includegraphics[width=1.0\linewidth]{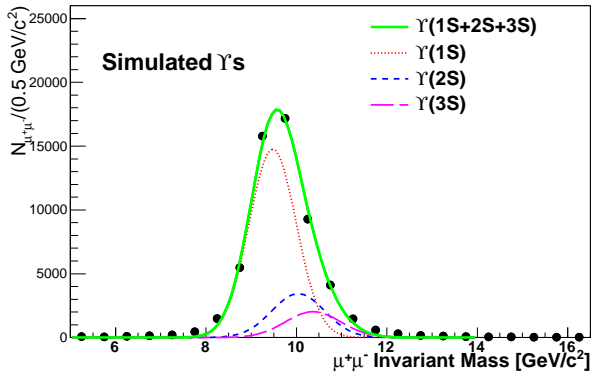}
  \caption{\label{fig:simupsilon} (color online) $\uall$ are generated 
  using {\sc pythia} 6.4 and run through the PHENIX {\sc geant}3 detector simulation.
  Each \ups has mass resolution of about 0.6 GeV$/c^{2}$. The line shape used for the fit function, Eq. (\ref{eq:fittingforupsilon})
  is composed of three Gaussians for the three \ups states.
  }
\end{figure}

Since the PHENIX muon-arm mass resolution is not good enough to resolve the 
three states of the \ups as shown in Fig.~\ref{fig:simupsilon}, we use 
results from two experiments at different collision energies and at 
different colliding systems to obtain an estimate of the relative ratio of 
the three \ups states, for the purpose of getting a distribution of line 
shape versus mass. The first is E605~\cite{e605}, a $p$ + A fixed target 
experiment at \sqsn=38.8~GeV and the second is CDF~\cite{cdf}, a collider 
experiment with $p+\overline{p}$ collisions at \sqs=1.8~TeV. These two 
experiments measured almost the same balance of the three \ups states 
although their energies and collision types are quite different 
(Table~\ref{table:Upsilon_composition}).

\begin{table}[!h]
\caption{Relative strength of the three \ups states at CDF~\cite{cdf} and 
from FNAL E605~\cite{e605}.}
\label{table:Upsilon_composition}
\begin{ruledtabular}
\begin{footnotesize}
\begin{tabular*}{\linewidth}{@{\extracolsep{\fill}}ccccc}
\ups states & Mass & Branching ratio                           & $p+\overline{p}$~\cite{cdf} & $p$ + A~\cite{e605}\\
                  & (GeV/$c^{2}$)& $\Upsilon \rightarrow \mu^{+}\mu^{-}$  & 1.8 TeV   & 38.8 GeV \\
\hline
$\uone$ & 9.46 & 2.48\% & 73$\%$ & 72$\%$ \\
$\utwo$ & 10.02 & 1.93\% & 17$\%$ & 19$\%$ \\
$\uthree$ & 10.36 & 2.29\% & 10$\%$ &  9$\%$ \\
\end{tabular*}
\end{footnotesize}
\end{ruledtabular}
\end{table}

As the energy of the measurement reported here is between those of the other 
measurements, we assume here that the composition of the three \ups states 
follows the ratio from CDF and assign a systematic uncertainty by varying 
the relative strength of the $\uone$ over $0.73 \pm 0.10$, with the 
fractions for the $\utwo$ and $\uthree$ changing accordingly. This 
uncertainty also accounts for the possibility that the 2S and 3S states 
might be suppressed more strongly than the 1S in \dau minimum-bias 
collisions, since it allows for a 30\% reduction in the 2S and a 50\% 
reduction in the 3S.

\begin{figure}[!h]
  \includegraphics[width=1.0\linewidth]{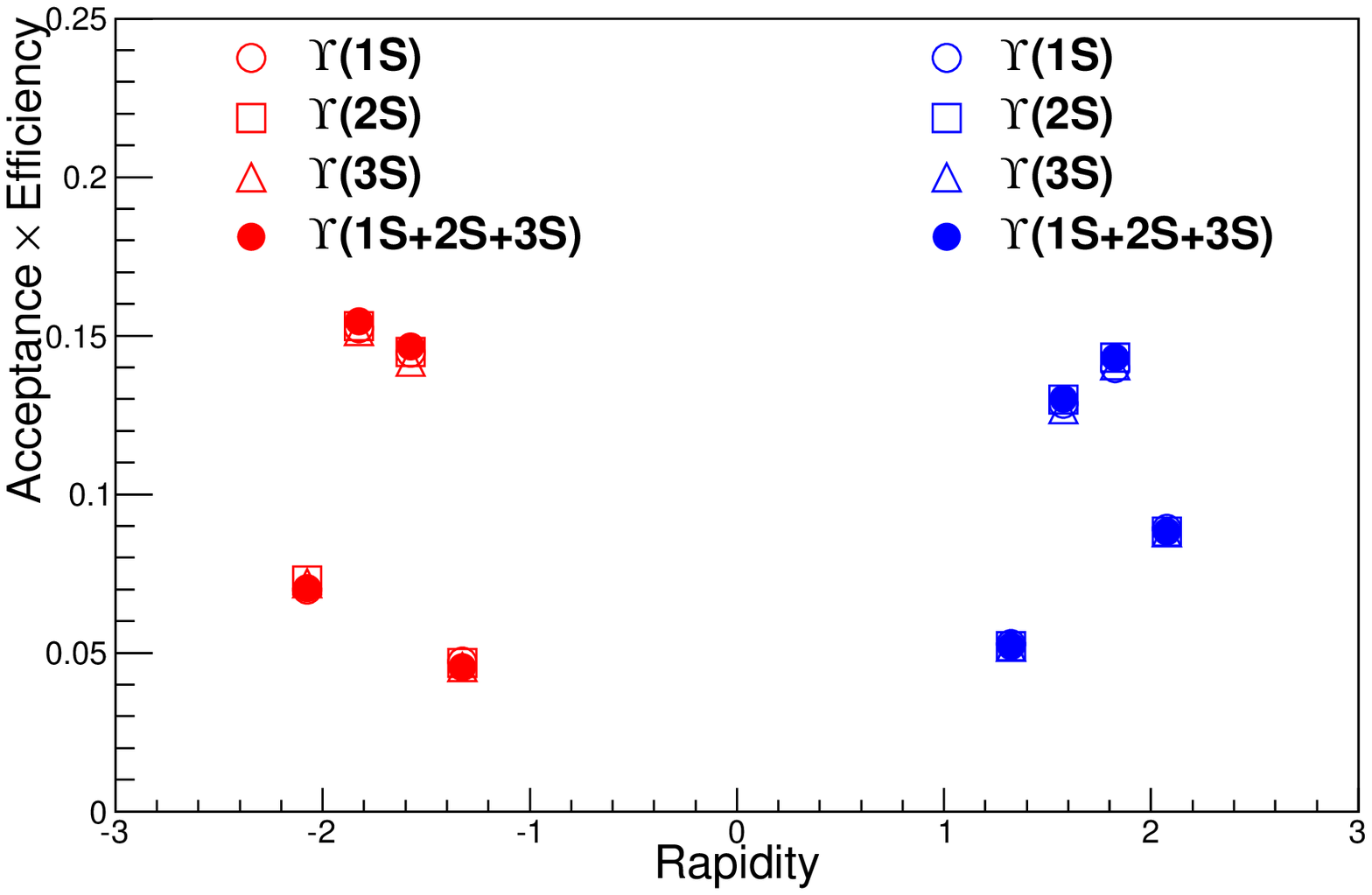}
  \caption{\label{fig:acceff} (color online) Acceptance$\times$Efficiency of each 
  \ups state and for the sum of the three states, $\uall$.
  The values for each state and for the sum of the three states are very similar.
  The $\uall$ is comprised of the $\uone$, $\utwo$, and 
  $\uthree$, following the abundances of 73$\%$ : 17$\%$ : 10$\%$ from the CDF experiment~\cite{cdf}.
  See also Table~\ref{table:Upsilon_composition}.}
\end{figure}

\ups simulations are performed for the three \ups states in order to 
estimate the effective \ups mass resolution and peak position for the real 
detector as well as to determine the acceptance-times-efficiency correction. 
To obtain these estimates, we generate the three \ups states with {\sc 
pythia} 6.4 and then process the generated \ups events through the PHENIX 
{\sc geant}3 simulation to make events with hits in the detectors. These 
simulated events are then embedded into real events to reflect the 
multiplicity environment of the data and are reconstructed with identical 
code to that used to reconstruct the data - including resolution smearing 
effects. The sum of three Gaussian functions is fit to the \ups-mass 
distribution. The results from the fit provide an estimate of the widths and 
the means of masses for the three \ups states. The resulting shape, as shown 
in Fig.~\ref{fig:simupsilon}, is then implemented for the fit function, Eq. 
(\ref{eq:fittingforupsilon}) used to extract the yields from the data.

Unlike the $J/\psi$, where the mass resolution is predominantly determined 
by effects from multiple scattering in the absorber preceding the 
muon-tracking volume, the higher momentum muons from the \ups experience 
less multiple scattering and less bending in the magnetic field; so the 
position resolution in the tracking volume becomes more important. To 
evaluate this, an additional systematic is obtained by allowing the mass 
resolutions of the three states to vary by $\pm$100 MeV/$c^{2}$ from their 
nominal, simulation determined, values.

With the simulated $\uall$ events, we also calculate the acceptance times 
efficiency ($A \epsilon_{\Upsilon}$) by dividing the reconstructed \ups 
yields by the {\sc pythia} generated \ups yields. Figure~\ref{fig:acceff} 
shows $A \epsilon_{\Upsilon}$ as a function of rapidity. $A 
\epsilon_{\Upsilon}$ of the summed $\uall$ and of each \ups states 
separately are quite similar to each other as shown. In this analysis, two 
inclusive rapidity bins are used, one for the positive rapidity and one for 
the negative rapidity. The values in \dau collisions from 2008 are 0.0950 
$\pm$ 0.0004 and 0.0980 $\pm$ 0.0004 for positive and negative rapidity, 
respectively. For \pp collisions, in the rapidity same order, from 2006 they 
are 0.1132 $\pm$ 0.0007 and 0.1096 $\pm$ 0.0007; and from 2009, they are 
0.1164 $\pm$ 0.0007 and 0.0907 $\pm$ 0.0007.

\begin{table*}[t!]
\caption{Parameter settings for the fits to the data and extraction of the
\upss [Eq. (\ref{eq:fittingforupsilon})].
See text for the details.}
\label{table:fitting}
\begin{ruledtabular}
\begin{tabular*}{\linewidth}{@{\extracolsep{\fill}}ccc}
Parameter & Fitting Parameter & Setting \\
\hline
$p_0$ & Yield of Drell-Yan process  & Fixed by NLO calculation \\
$p_1$ & Slope of Drell-Yan process  & Fixed by NLO calculation \\
$p_2$ & Slope of Drell-Yan process  & Fixed by NLO calculation \\
$p_3$ & Slope of Drell-Yan process  & Fixed by NLO calculation \\
\\
$p_4$ & Yield of Charm/Beauty correlations    & Set free \\ 
$p_5$ & Slope of Beauty correlation     & Fixed by {\sc pythia}/{\sc geant} simulation \\
$p_6$ & Relative ratio of Charm/Beauty correlations & Fixed PHENIX dilepton measurement~\cite{Adare:dilepton} \\
$p_7$ & Slope of Charm correlation     & Fixed by {\sc pythia}/{\sc geant} simulation \\
\\
$p_8$ & Yield of $\uall$         & Set free \\
$p_9$ & Mean value of $\uone$    & Fixed by {\sc pythia}/{\sc geant} simulation \\
$p_{10}$ & Resolution of $\uone$   & Fixed by {\sc pythia}/{\sc geant} simulation \\
$p_{11}$ & Relative ratio of $\utwo$  & Fixed by CDF experiment~\cite{cdf} \\
$p_{12}$ & Mean value of $\utwo$   & Fixed by {\sc pythia}/{\sc geant} simulation \\
$p_{13}$ & Resolution of $\utwo$   & Fixed by {\sc pythia}/{\sc geant} simulation \\
$p_{14}$ & Relative ratio of $\uthree$  & Fixed by CDF experiment~\cite{cdf} \\
$p_{15}$ & Mean value of $\uthree$   & Fixed by {\sc pythia}/{\sc geant} simulation \\
$p_{16}$ & Resolution of $\uthree$   & Fixed by {\sc pythia}/{\sc geant} simulation \\
\end{tabular*}
\end{ruledtabular}
\end{table*}

\begin{table*}[t!]
\caption{Systematic uncertainties for each source and for each collision type (see the text for details). 
        Type A represents a point-to-point uncorrelated systematic uncertainty. 
        Type B represents a common systematic uncertainty between points at different rapidity. Type C is a global uncertainty.}
\label{table:systematic_uncertainty}
\begin{ruledtabular}
\begin{tabular*}{\linewidth}{@{\extracolsep{\fill}}cccccccccc}
Systematic uncertainty sources & \pp South & \pp North & \dau South & \dau North & Type \\
  			& (backward rapidity)& (forward rapidity) & (backward rapidity)& (forward rapidity) & \\
\hline
Relative ratio of $\uall$    & 2.6 $\%$ & 0.8 $\%$ & 0.9 $\%$ & 2.6 $\%$ & A\\
Relative ratio of Bottom/Charm         		 & 0.3 $\%$ & 0.1 $\%$ & 0.4 $\%$ & 0.3 $\%$ & A\\
\ups mass resolution                          & 6.4 $\%$ & 6.7 $\%$ & 7.9 $\%$ & 8.3 $\%$ & B\\
NLO DY model        				 & 7.7 $\%$ & 6.9 $\%$ & 4.0 $\%$ & 4.4 $\%$ & B\\
NLO DY renormalization/factorization    		 & 0.9 $\%$ & 0.9 $\%$ & 7.7 $\%$ & 2.2 $\%$ & B\\
Open Bottom random correlation      		 & 7.2 $\%$ & 4.0 $\%$ & 9.2 $\%$ & 6.3 $\%$ & B\\
Combinatorial background normalization           & 0.6 $\%$ & 0.7 $\%$ & 4.0 $\%$ & 2.3 $\%$ & B\\
Combinatorial background estimation              & 1.7 $\%$ & 2.3 $\%$ & 0.0 $\%$ & 1.1 $\%$ & B\\
methods of like-sign and mixed events                    &          &          &          &          & \\   
MuID efficiency         					 & 4.0 $\%$ & 4.0 $\%$ & 4.0 $\%$ & 4.0 $\%$ & B\\
MuTr efficiency    					 & 2.0 $\%$ & 2.0 $\%$ & 2.0 $\%$ & 2.0 $\%$ & B\\
BBC efficiency           				 & 10.1 $\%$ & 10.1 $\%$ & 5.3 $\%$ & 5.3 $\%$ & C\\
\end{tabular*}
\end{ruledtabular}
\end{table*}

\subsection{Data evaluation and \ups extraction \label{sec:fitting}}

\begin{figure}[!h]
  \includegraphics[width=1.0\linewidth]{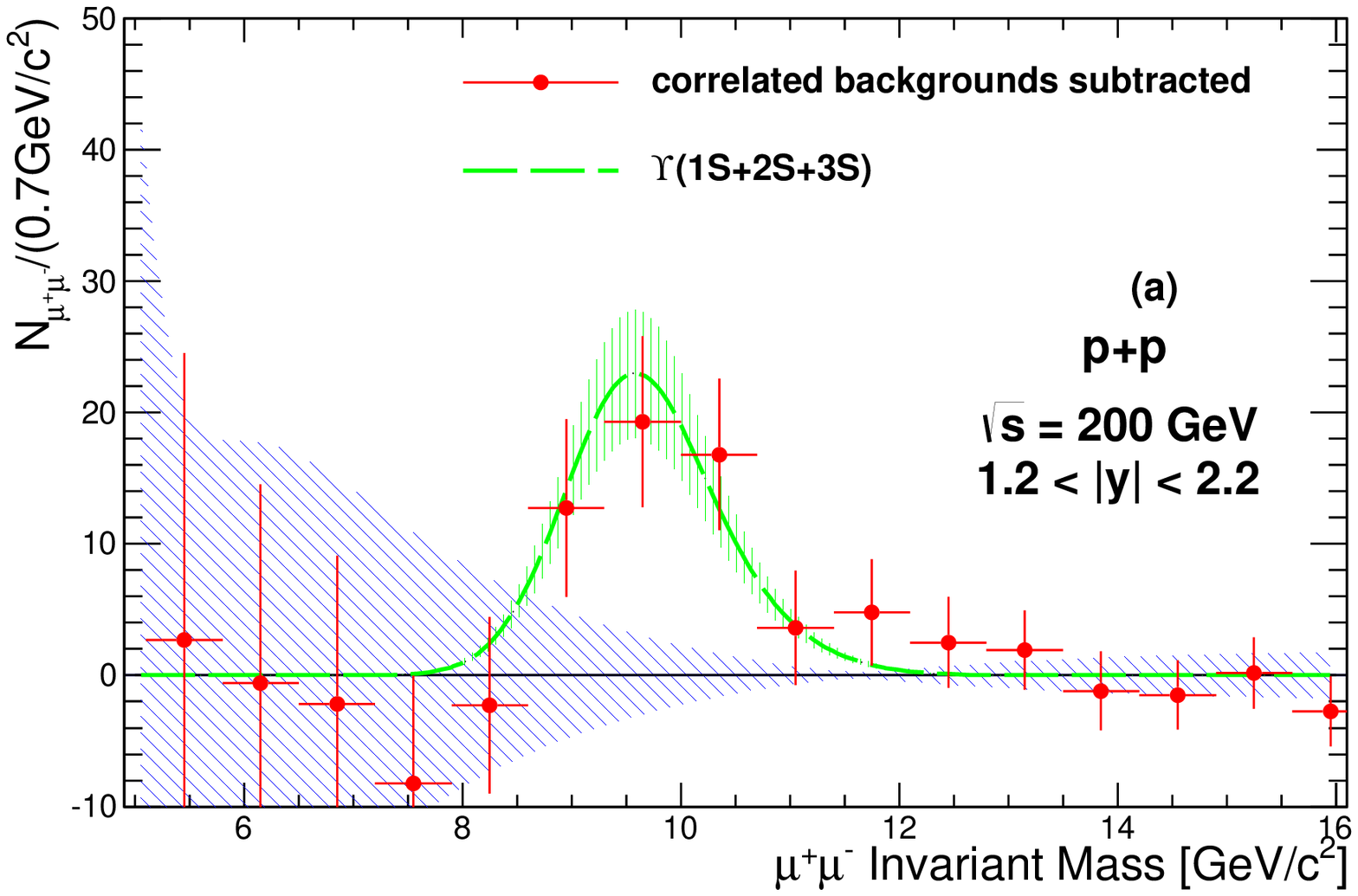}
  \includegraphics[width=1.0\linewidth]{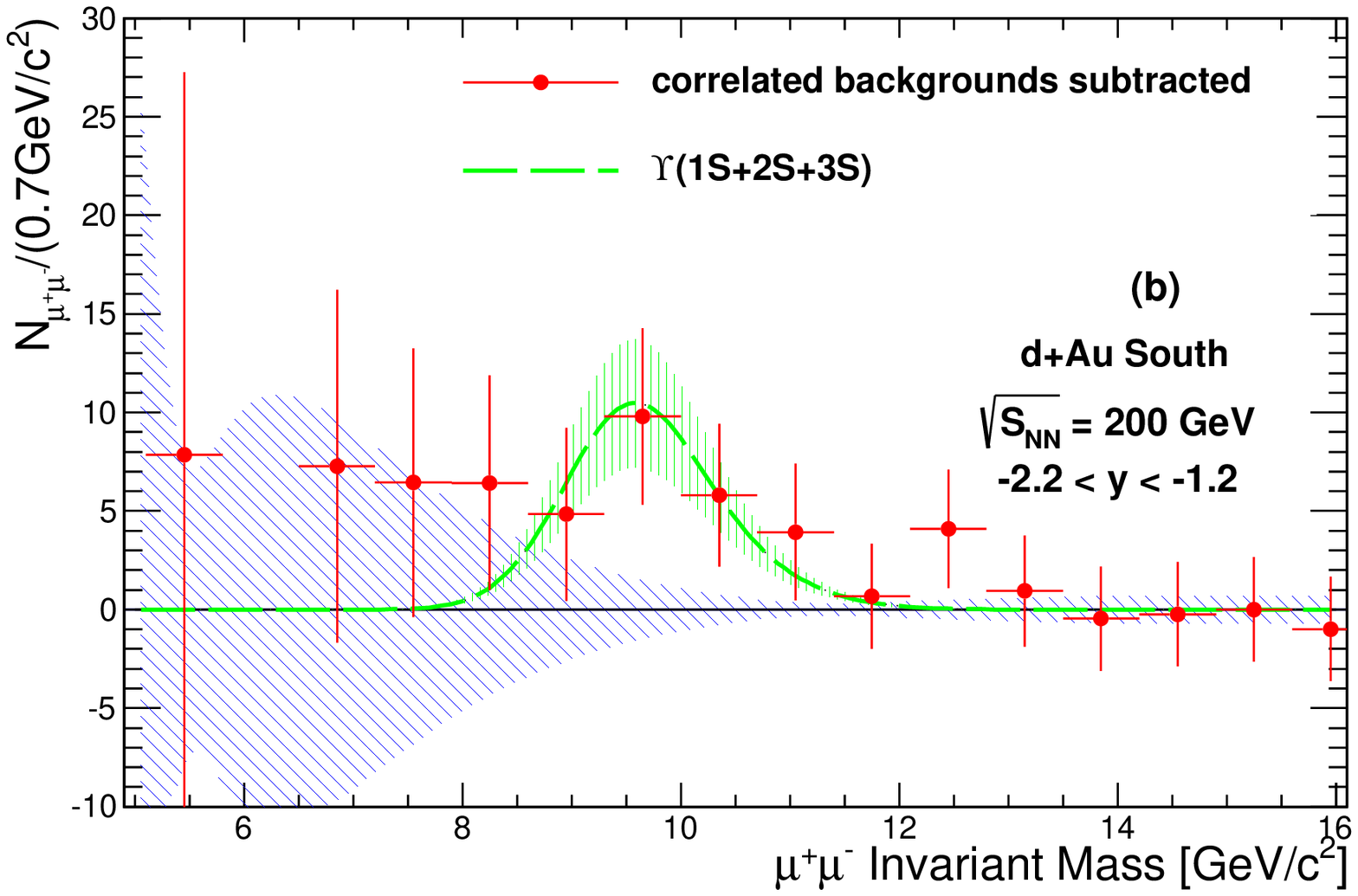}
  \includegraphics[width=1.0\linewidth]{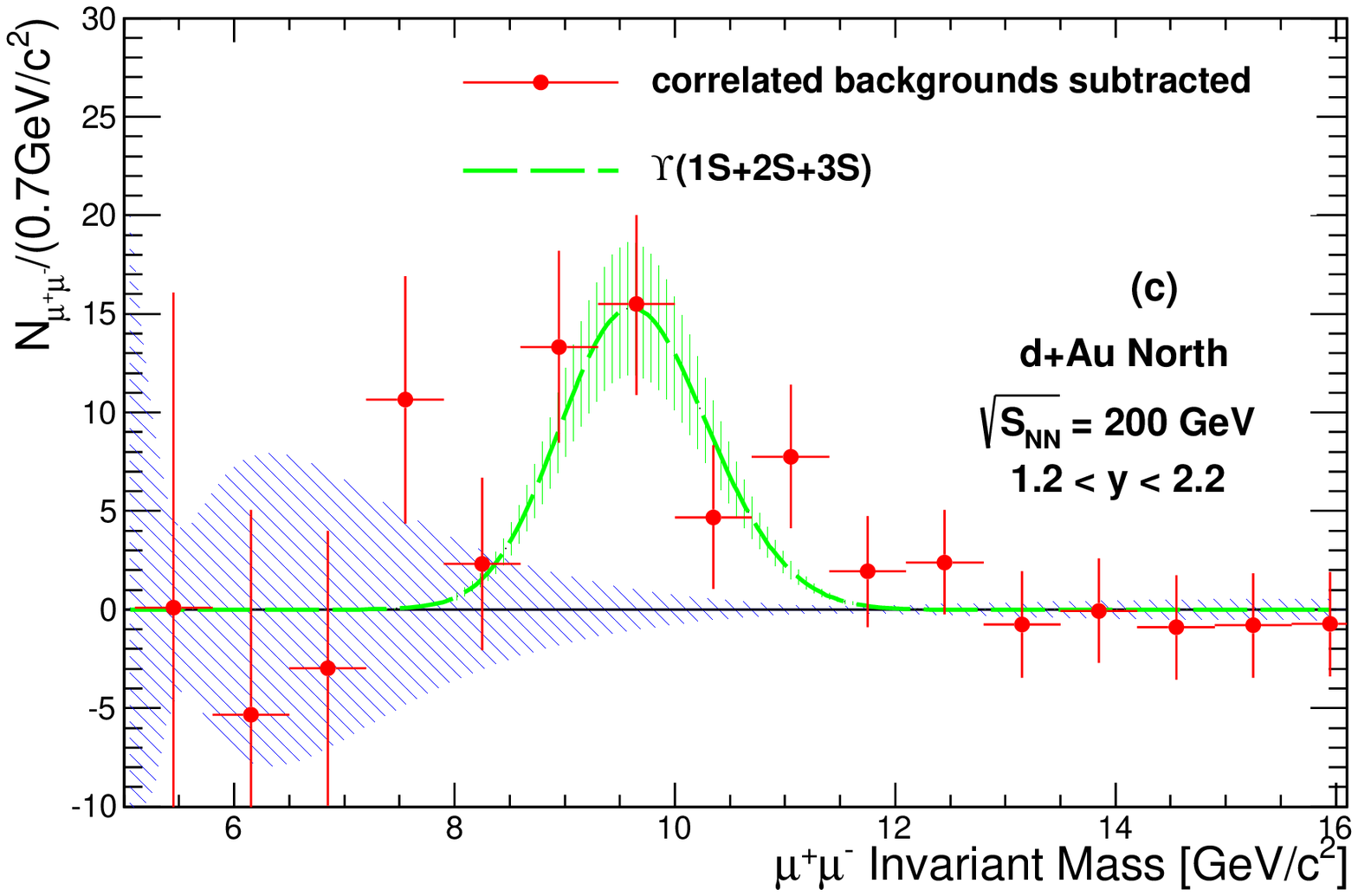}
  \caption{\label{fig:upsilondist} (color online)
  The mass distributions are drawn after subtraction of all the 
  correlated backgrounds and Drell-Yan process. The fitted peak curves represent only $\uall$.  
  The shaded bands (green vertical shading) around the fitted peak curves represent fitting uncertainties
  and those around zero yield (blue slanted shading) represent systematic uncertainties
  from the different assumptions for NLO Drell-Yan models, {\sc pythia} and random correlations of charm/bottom. 
  }
\end{figure}

We extract \upss from the data using the estimated correlated backgrounds 
and the \upss as described in Sections~\ref{sec:drellyan}, 
\ref{sec:opencorrelation}, and \ref{sec:upsilon}.  The fit function used for 
the mass distribution is shown in Eq. (\ref{eq:fittingforupsilon}). In 
addition to the $\uall$ signal, the function includes contributions from the 
Drell-Yan process and from correlated open bottom/charm pairs.

\begin{widetext}
\begin{eqnarray}
F(m) = p_{0}{\rm exp}(p_{1}m+p_{2}m^{2}+p_{3}m^{3})+p_{4}[(1-p_{6}){\rm exp}(p_{5}m)+p_{6}{\rm exp}(p_{7}m)]+ \nonumber \\ 
p_{8}[(1-p_{11}-p_{14}){\rm exp}(-0.5A^2)+p_{11}{\rm exp}(-0.5B^2)+p_{14}{\rm exp}(-0.5C^2)]
\label{eq:fittingforupsilon}
\end{eqnarray}
\end{widetext}

\noindent
where $m$ is the invariant mass of the dimuon, $A$ $= (m-p_{9})/p_{10}$, 
$B$=$(m-p_{12})/p_{13}$, and $C$=$(m-p_{15})/p_{16}$.
The parameters $p_{0}$ to $p_{3}$ are for the NLO Drell-Yan process and
are fixed by the NLO calculations and the PHENIX {\sc geant}3 simulation, 
as discussed in Section~\ref{sec:drellyan}.

The parameters $p_{4}$ to $p_{7}$ are for the contribution of the open 
bottom and open charm correlated pairs. The relative ratio of bottom to 
charm yields, which is represented by $p_{6}$, is fixed from the PHENIX 
dilepton mass spectra study~\cite{Adare:dilepton}, and the shape is 
determined from the {\sc pythia} and the PHENIX {\sc geant}3 simulation, as 
described in Section~\ref{sec:opencorrelation}. The total yield from 
correlated bottom and charm, $p_{4}$, is allowed to vary in the fits to the 
data. The resulting contributions from correlated bottom and charm are then 
checked against those from the PHENIX dilepton measurements, which have 
bottom and charm cross sections of $\sigma_{c\overline{c}}$=544 
$\pm$~39~(stat) $\pm$~142~(syst) $\pm$~200~(model)~$\mu$b and 
$\sigma_{b\overline{b}}$=3.9 $\pm$~2.5~(stat) 
$^{+3}_{-2}$~(syst)~$\mu$b~\cite{Adare:dilepton}. For this check, we 
integrated our fitted correlated bottom and charm over the mass range 5 
GeV/$c^{2}$ to 16 GeV/$c^{2}$, and then added the contribution from 
unmeasured regions assuming the mass shape from the NLO calculation. This 
estimate of the contribution of bottom and charm is within a half sigma in 
the experimental uncertainties of the nominal cross sections. For \dau 
collisions, this estimate is still within a half sigma when we scale the 
nominal cross section by the number of participants (2$\times$197) assuming 
no nuclear modification effects on the production.

The parameters $p_{8}$ to $p_{16}$ are for the contribution of the $\uall$. 
$A$, $B$, and $C$ represent the means of masses and widths of the three \ups 
states, as estimated using the PHENIX {\sc geant}3 simulation package and 
fixed for data - as described in Section~\ref{sec:upsilon}. The total yield 
of \ups, $p_{8}$, is allowed to vary in the fits to the data in order to 
extract the \ups signal.

The data are fit using a log-likelihood fitting method that adds the 
normalized combinatorial background to both the mass distribution and the 
fit function. This has the advantage that empty bins in the mass 
distribution, which result from statistical fluctuations of the background 
above the signal size and otherwise produce negative counts, are accounted 
for properly. The fitting quality is very good with $\chi^2$/dof of 9.0/16 
and 6.4/16 for the negative and positive rapidities of \pp collisions, 
respectively, and 14.6/16 and 9.5/16 for the negative and positive 
rapidities of \dau collisions, respectively. Systematic uncertainties from 
the NLO calculation, the assumed cross sections, and detector performances 
are explained in Sections~\ref{sec:drellyan}, \ref{sec:opencorrelation}, and 
\ref{sec:upsilon}, and are summarized in 
Table~\ref{table:systematic_uncertainty}. Figure~\ref{fig:upsilondist} (a), 
(b), and (c) show the $\uall$ mass distribution after subtracting off all 
correlated backgrounds, for \pp and \dau collision systems. We checked above 
the \ups-mass region, $>$ 11.5 GeV/$c^{2}$, for fitting reliability; the 
integral of the high-mass region is within the systematical uncertainties, 
which are drawn as shaded bands.

\section{Results and Discussion \label{sec:result}}

\begin{table}[hb!]
\caption{\ups invariant yields and cross sections of
\pp and \dau data sets are shown. The first uncertainty shown is the statistical,
and the second is the systematical.}
\label{table:invyield}
\begin{ruledtabular}
\begin{tabular*}{\linewidth}{@{\extracolsep{\fill}}cccccccccc}
Data set            & N$_{\rm MB}^{\rm BBC}$  & $B_{\mu^{+}\mu^{-}}$d$N_{\Upsilon}$/dy & $B_{\mu^{+}\mu^{-}}$ d$\sigma_{\Upsilon}$/dy\\
                    & ($\times 10^{11}$) &      ($\times 10^{-10}$)       & \\
\hline
\pp South     & 5.1        & 3.8 $\pm$ 1.1 $\pm$ 0.5 & 16.0 $\pm$ 4.6 $\pm$ 2.2 pb \\
\pp North     & 5.2        & 3.6 $\pm$ 1.0 $\pm$ 0.4 & 15.4 $\pm$ 4.3 $\pm$ 1.8 pb\\
\dau South      & 1.3        & 17.8 $\pm$ 5.5 $\pm$ 2.9 & 4.0 $\pm$ 1.2 $\pm$ 0.6 nb\\
\dau North      & 1.3        & 25.2 $\pm$ 5.6  $\pm$ 3.3 & 5.7 $\pm$ 1.3 $\pm$ 0.7 nb\\
\end{tabular*}
\end{ruledtabular}
\end{table}

The invariant yields of $\uall$ for each rapidity bin are calculated as,

\begin{equation}
B_{\mu^+\mu^-} \frac{dN_{\Upsilon}}{dy} = \frac{N_{\Upsilon} \cdot C}{N_{\rm MB} \cdot A \epsilon_{\Upsilon} \cdot \Delta{y}},
\label{eq:invy}
\end{equation}

\noindent
where the notation is the same as for Eq. (\ref{eq:dycalculation3}), except 
\ups is used instead of the Drell-Yan process. Table~\ref{table:invyield} 
shows calculated yields and cross sections for both arms and for both 
collision types.

\begin{figure}[]
  \includegraphics[width=1.0\linewidth]{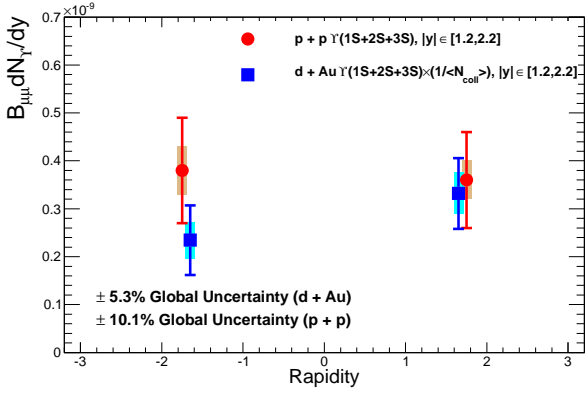}
  \caption{\label{fig:upsilonyield} (color online) Invariant yields, 
  $B_{\mu\mu} dN_{\Upsilon}/dy$, for \pp and \dau collisions are shown as a 
  function of rapidity. The solid error bars represent the statistical uncertainties.
  The boxes represent the systematic uncertainties.
  The global systematic uncertainties are quoted as text at the bottom.
  }
\end{figure}

Figure~\ref{fig:upsilonyield} shows the invariant \ups yields for \pp and 
\dau collisions. The nuclear modification factor, \rdau, can be obtained 
from the invariant yields. Eq.~(\ref{eq:upsilonrdau}) shows the relation 
between \rdau, the invariant yield, and $N_{\rm coll}$. The scale factor, 
$N_{\rm coll}$, makes \rdau one if the \ups yield for \dau collisions is 
equal to the \ups yield for \pp collisions times the number of binary 
collisions in \dau collisions, i.e. \rdau=1 if there are no nuclear 
modification effects.

\begin{figure}[]
  \includegraphics[width=1.0\linewidth]{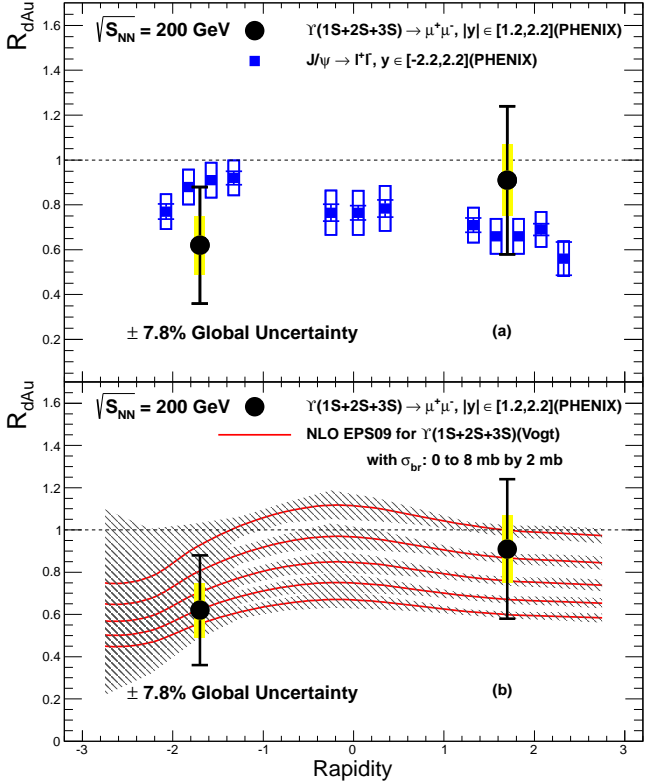}
  \caption{\label{fig:rdau} (color online)
  Nuclear modification factors, \rdau, are shown as a function of rapidity.
  For comparison, the upper panel (a) shows \rdau for the $J/\psi$~\cite{jpsirdau2011} as well as for the \ups.
  The solid error bars represent the statistical uncertainties and the boxes 
  represent the systematic uncertainties. The global systematic uncertainty is 
  quoted as text at the bottom. The lower panel (b) shows theoretical
  predictions of nuclear modification based on a NLO EPS09 combined with a breakup
  cross section, with $\sigma_{br}$=0 to 8 mb in 2 mb steps from top to bottom.
  See text for the details.}
\end{figure}

As seen in Fig.~\ref{fig:rdau}, at forward rapidity \ups production shows no 
significant suppression with an \rdau of 0.91 $\pm$ 0.33 (stat) $\pm$ 0.16 
(syst); while at backward rapidity the suppression of the \ups is 
approximately one sigma (of the experimental uncertainty) below one with an 
\rdau of 0.62 $\pm$ 0.26 (stat) $\pm$ 0.13 (syst). Figure~\ref{fig:rdau} (a) 
shows a comparison to previous results from PHENIX for \rdau of the 
$J/\psi$. The $J/\psi$ results show a larger suppression at forward than at 
backward rapidity, a trend that cannot be confirmed or denied for the \ups 
given the large uncertainties of the measurements presented here.

A NLO calculation with EPS09 shadowing and a breakup cross 
section~\cite{ramonardau} predicts modest suppression at backward rapidity, 
but no shadowing at forward rapidity; although there could be suppression by 
a breakup cross section, as seen in the red lines in Fig.~\ref{fig:rdau} 
(b). The rapidity dependence of this NLO calculation appears to be 
consistent with the trend between our backward- and forward-rapidity 
measurements. At both backward and forward rapidities, the large 
uncertainties of the measurements do not give a significant constraint on 
the breakup cross section within the context of the NLO models. It will also 
be of interest to compare other models that include the effects of 
initial-state parton energy loss or of gluon saturation to this \ups data.

\begin{figure}[]
  \includegraphics[width=1.0\linewidth]{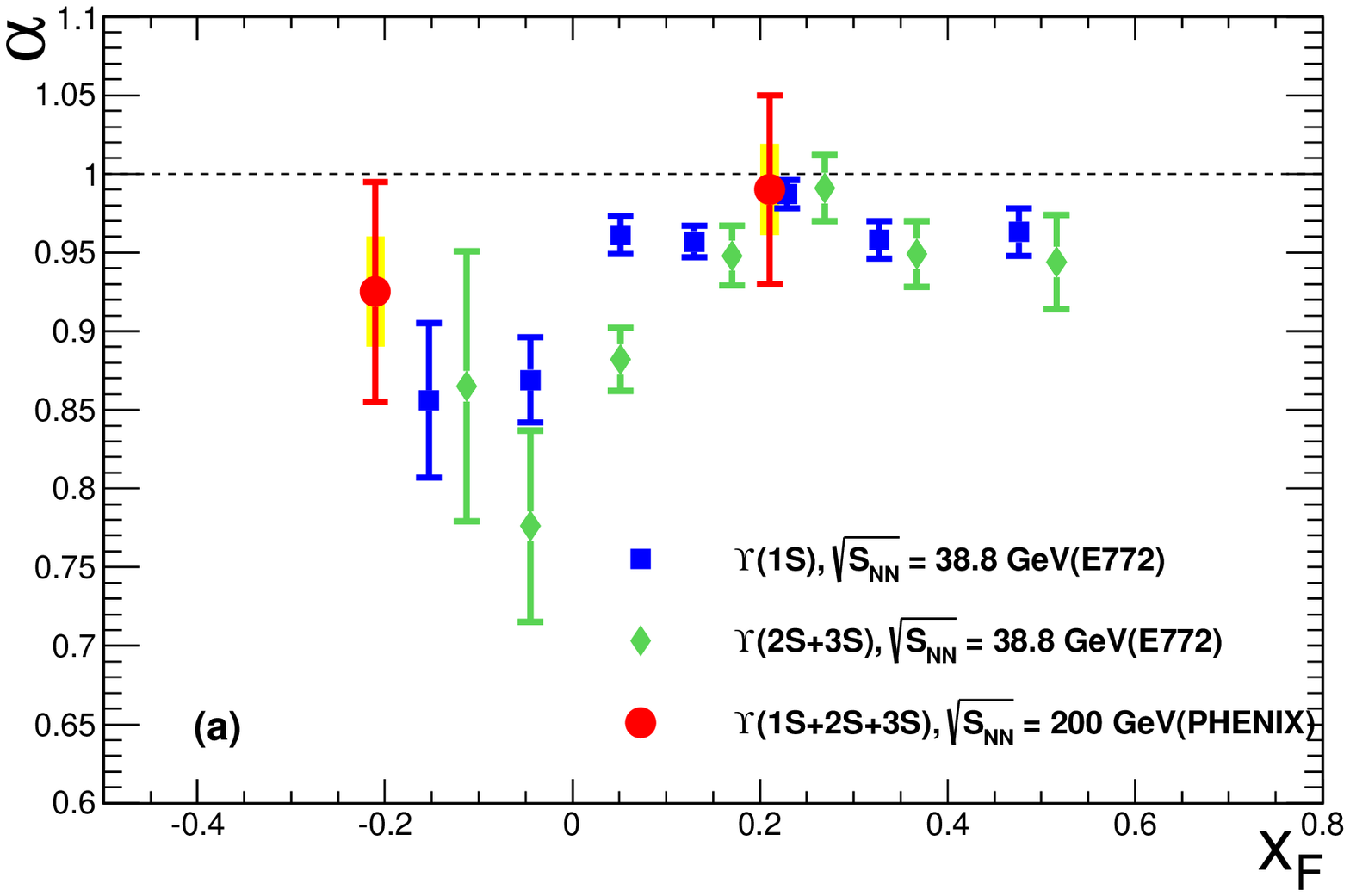}
  \includegraphics[width=1.0\linewidth]{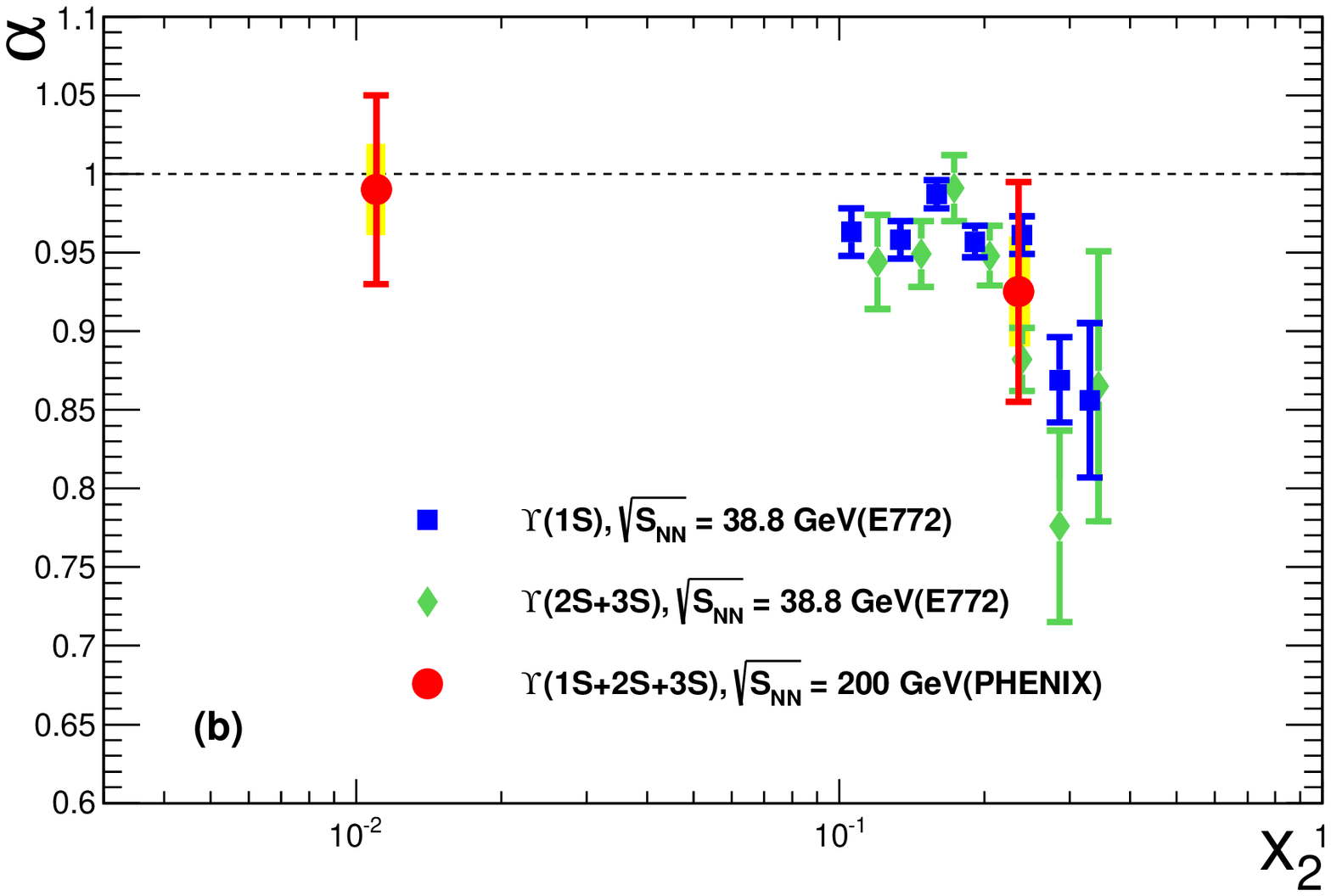}
  \caption{\label{fig:lowerenergy} (color online) $\alpha$ versus $x_F$ and $x_2$.
  The square and diamond shaped points are from the E772 experiment
  where an 800 GeV proton beam collides with fixed targets of
  $^2$H, C, Ca, Fe and W, corresponding to \sqsn=38.8 GeV. The round 
  points are from this analysis. The solid error bars represent the statistical uncertainties and
  the boxes represent the systematic uncertainties.}
\end{figure}

We can parameterize the nuclear dependence of \ups production as 
$\sigma_{\Upsilon}^{\dau}=\sigma_{\Upsilon}^{\pp} \times (2A_{Au})^{\alpha}$ 
for \dau collisions, where $A_{Au}$ represents the number of nucleons in the 
gold nucleus. As for \rdau, if there are no nuclear effects then $\alpha$ 
would be one. Previously, E772, which was at \sqsn=38.8 GeV, showed a large 
decrease in $\alpha$ at $x_{F}$ $<$ 0. The PHENIX backward-rapidity covers 
-0.42 $\leq$ $x_{F}$ $\leq$ -0.14, where $x_{F}=x_{1} - x_{2}$ and $x_{1}$ 
is the momentum fraction of the gluon in deuteron. The backward-rapidity 
($\left<x_F\right>$ $\sim$ -0.2, $\left<x_2\right>$ $\sim$ 2 $\times$ 
$10^{-1}$) PHENIX measurements obtain $\alpha_{\uall}=$ 0.925 
$\pm$~0.070~(stat) $\pm$~0.035~(syst) and at forward rapidity 
($\left<x_F\right>$ $\sim$ 0.2, $\left<x_2\right>$ $\sim$ 1 $\times$ 
$10^{-2}$) $\alpha_{\uall}=$ 0.990 $\pm$~0.060~(stat) $\pm$~0.029~(syst).

Figure~\ref{fig:lowerenergy} shows $\alpha$ versus $x_{F}$ and versus 
$x_{2}$ from the E772 data and from our data. The suppression levels of 
$\uall$ in PHENIX are consistent with those from E772 within uncertainties.

For our \dau measurements, we can also calculate the ratio of the \ups 
yield, $dN_{\Upsilon}^{\dau}/dy$, between backward and forward rapidities as 
a test of the nPDF. This ratio, $dN_{\Upsilon}^{\dau}/dy|_{-2.2 < y < -1.2}$ 
/ $dN_{\Upsilon}^{\dau}/dy|_{1.2 < y < 2.2}$, shows some suppression at 
backward rapidity relative to forward rapidity, with a value of 0.71 $\pm$ 
0.27 (stat); but the effect is not very significant due to the large 
uncertainty.

\section{Summary \& Conclusion \label{sec:summary}}

In summary, we have presented the first yields, cross sections, and nuclear 
dependences for \ups production in \sqsn=200~GeV \dau and \pp collisions for 
two rapidity bins. At backward rapidity, $\uall$ yields are measured to be 
suppressed by approximately one sigma of the experimental uncertainty below 
one. The rapidity dependence of the observed \ups suppression at forward and 
backward rapidities are compatible with lower energy results and a NLO 
theoretical calculation. Comparison to the theoretical calculation for a 
model that includes EPS09 shadowing and a breakup cross section does not 
result in any definitive constraint on the breakup cross section given the 
large experimental uncertainties. Future comparisons to gluon saturation 
models and to models including initial-state energy loss would also be of 
interest.


\section*{Acknowledgements}   

We thank the staff of the Collider-Accelerator and Physics
Departments at Brookhaven National Laboratory and the staff of
the other PHENIX participating institutions for their vital
contributions.  
We also thank Ramona Vogt, Ivan Vitev, and Rishi Sharma
for useful discussions and theoretical calculations.
We acknowledge support from the 
Office of Nuclear Physics in the
Office of Science of the Department of Energy, 
the National Science Foundation, 
a sponsored research grant from Renaissance Technologies LLC,
Abilene Christian University Research Council, 
Research Foundation of SUNY, and 
Dean of the College of Arts and Sciences, Vanderbilt University 
(U.S.A),
Ministry of Education, Culture, Sports, Science, and Technology
and the Japan Society for the Promotion of Science (Japan),
Conselho Nacional de Desenvolvimento Cient\'{\i}fico e
Tecnol{\'o}gico and Funda\c c{\~a}o de Amparo {\`a} Pesquisa do
Estado de S{\~a}o Paulo (Brazil),
Natural Science Foundation of China (P.~R.~China),
Ministry of Education, Youth and Sports (Czech Republic),
Centre National de la Recherche Scientifique, Commissariat
{\`a} l'{\'E}nergie Atomique, and Institut National de Physique
Nucl{\'e}aire et de Physique des Particules (France),
Bundesministerium f\"ur Bildung und Forschung, Deutscher
Akademischer Austausch Dienst, and Alexander von Humboldt Stiftung (Germany),
Hungarian National Science Fund, OTKA (Hungary), 
Department of Atomic Energy and Department of Science and Technology (India), 
Israel Science Foundation (Israel), 
National Research Foundation and WCU program of the 
Ministry Education Science and Technology (Korea),
Ministry of Education and Science, Russian Academy of Sciences,
Federal Agency of Atomic Energy (Russia),
VR and Wallenberg Foundation (Sweden), 
the U.S. Civilian Research and Development Foundation for the
Independent States of the Former Soviet Union, 
the Hungarian American Enterprise Scholarship Fund,
and the US-Israel Binational Science Foundation.


\end{document}